\documentclass[10pt,a4paper]{article}

\usepackage[british]{babel}

\usepackage[a4paper,top=2cm,bottom=2cm,left=2.5cm,right=2.5cm,marginparwidth=1.75cm]{geometry}

\usepackage[square,numbers,sort&compress]{natbib}
\usepackage{amsmath}
\usepackage{amssymb}
\usepackage{graphicx}
\usepackage[colorlinks=true, allcolors=blue]{hyperref}
\usepackage{hyperref}
\usepackage[title]{appendix}
\usepackage{mathrsfs}
\usepackage{amsfonts}
\usepackage{booktabs} 
\usepackage[format=plain,
            labelfont={bf,sl}]{caption}
\usepackage{threeparttable} 
\usepackage{algorithm}
\usepackage{algorithmicx}
\usepackage{algpseudocode}
\usepackage{listings}
\usepackage{enumitem}
\usepackage{chngcntr}
\usepackage{booktabs}
\usepackage{lipsum}
\usepackage{subcaption}
\usepackage{authblk}
\usepackage[T1]{fontenc}    
\usepackage{csquotes}       
\usepackage{diagbox}
\usepackage{xcolor}
\usepackage{textcomp}
\usepackage{gensymb}

\usepackage{siunitx}
\usepackage{hhline}

\usepackage{setspace}
\usepackage{titlesec}
\titleformat{\section} 
  {\normalfont\Large\bfseries}{\thesection.}{1em}{}
  
\usepackage{lineno} 

\rightlinenumbers 

\usepackage{float}   
\usepackage{caption} 



\title{Quantum transport in SnTe nanowire devices}

\author[1]{Femke J. Witmans}
\author[2]{Mathijs G. C. Mientjes}
\author[1]{Maarten J. G. Kamphuis}
\author[2]{Vince van de Sande}
\author[2]{Xin Guan}
\author[3]{Hans Bolten}
\author[2,3]{Marcel A. Verheijen}
\author[1]{Chuan Li}
\author[1]{Joost Ridderbos}
\author[2]{Erik P. A. M. Bakkers}
\author[1]{Alexander Brinkman}
\author[1,*]{Floris A. Zwanenburg}
\affil[1]{\small MESA+ Institute for Nanotechnology, University of Twente, P.O. Box 217, 7500 AE Enschede, The Netherlands}
\affil[2]{\small Department of Applied Physics, Eindhoven University of Technology, P.O Box 513, 5600 MB Eindhoven, The Netherlands}
\affil[3]{Eurofins Materials Science Eindhoven, 5656 AE Eindhoven, The Netherlands}
\affil[*]{Corresponding author: \texttt{f.a.zwanenburg@utwente.nl}}

\date{}  

\begin{document}
\maketitle

\begin{abstract}
We report on a variety of quantum transport experiments in SnTe nanowire devices. Research on these particular nanowire devices is relevant because of their topological properties and their potential to distinguish surface states owing to their high surface-to-volume ratio that suppresses the bulk contribution to the conductance. We observe a low-resistance and a high-resistance regime. The highly resistive devices display semiconducting and quantum dot behavior caused by microscopic differences in the fabrication, while devices with low resistance show partial superconductivity when in a hybrid superconductor-nanowire configuration or Fabry-Pérot oscillations. The latter suggests quantum interference in a ballistic transport channel, attributed to the 2D surface states in SnTe. The wide variety of quantum transport phenomena demonstrate SnTe nanowires as a promising platform for diverse follow-up experiments and novel device architectures, including the exploration of topological superconductivity and the development of low-energy spintronic devices.    

\end{abstract}


\section{Introduction}\label{introduction}
Since the theoretical prediction in 2005 \cite{Kane2005}, interest in topological insulators (TIs) has grown significantly \cite{ReviewTI2011}. TIs feature an insulating bulk with conducting surface states when interfaced with a topologically trivial material \cite{FuKaneMele2007}. On the conducting surfaces, the spin is locked perpendicular to the momentum, which make the surface states immune to non-magnetic perturbations \cite{Fu2007, Konig2007}. This class of materials serves as a compelling platform for exploring unique quantum mechanical effects. Such effects form the basis for potential applications such as low-dissipation transport, spintronics, and quantum computation \cite{Moore2010}. Topological crystalline insulators (TCIs) are a specific type of TI in which the topological nature of the electronic structure arises from crystal symmetry \cite{Fu2011}. The first material predicted to be a TCI was SnTe \cite{Hseih2012}, with metallic surface states on the high-symmetry crystal surfaces, which are symmetric around the \{110\} mirror planes. Note that already in 1987 the Pb\textsubscript{1-x}Sn\textsubscript{x}Te system was predicted to have a band inversion described by the Dirac Hamiltonian \cite{PANKRATOV198793}.

Since the discovery of SnTe as a TCI, it has attracted much attention for theoretical research, especially because of the relatively simple rock salt crystal structure and stoichiometry \cite{Hseih2012,Bauer2013}. However, in experiments one of the disadvantages of SnTe is the dominant transport of bulk carriers due to the spontaneous formation of Sn vacancies in the material \cite{Snvacancy2014}, which results in the Fermi level being situated deeply in the valence band. Transport experiments on SnTe are limited, presumably due to the dominant influence of the 3D bulk carriers on transport properties. Previous studies have shown signatures of surface states, such as beating patterns in Shubnikov-de Haas oscillations \cite{dybko_experimental_2017,costa_investigation_2021,okazaki_shubnikov-haas_2018,taskin_topological_2014,Safdar2013} and weak antilocalization \cite{liu_cr_2024,assaf_quantum_2014,zou_revealing_2019,yan_structure_2020,akiyama_two-dimensional_2016,de_castro_weak_2022,akiyama_weak_2014,li_weak_2022,albright_weak_2021}, although in all cases accompanied by transport through the bulk. In contrast, the lack of surface state transport is reported on as well \cite{khaliq_low_2022,klett_proximity-induced_2018,liu_synthesis_2021}, emphasizing the difficulty of reducing the influence of the bulk carriers. This calls for investigating SnTe nanowires to increase the surface-to-volume ratio, leading to an enhancement of the surface state contribution. Very little experimental work has been published on narrow SnTe nanowires \cite{Safdar2013, Trimble2021}. Safdar et al. reported \cite{Safdar2013} on magnetotransport measurements including Aharanov-Bohm and Shubnikov-de Haas oscillations in SnTe nanowires. Trimble et al. \cite{Trimble2021} reported proximity-induced superconductivity and the breaking of time-reversal symmetry, a phenomenon that can partially be explained by the ferroelectric phase transition previously observed in SnTe \cite{liu_synthesis_2021,baral_growth_2019,shen_synthesis_2014}. The studies do not report any tunability in the chemical potential, neither in the bulk nor surface states. Thus, in this work, we use SnTe nanowires, grown similarly as in Ref. \cite{Mientjes2024}. Compared to previous reports, our wires are notably thin with a maximum thickness of 60~nm, compared to nanowire thicknesses of 160–300~nm used in other transport studies. Here we report three quantum transport phenomena in SnTe nanowire devices: semiconducting behavior such as (partial) depletion of the nanowire channel, the onset of superconductivity in a hybrid superconductor-nanowire device, and the establishment of Fabry-Pérot oscillations in the nanowire channel. This shows that a wide spectrum of experiments is possible in these nanowires.

\section{Results and discussion}
\subsection{Structural characterization}\label{sec:morph}
\begin{figure}[!ht]
\centering
\includegraphics[width=\linewidth]{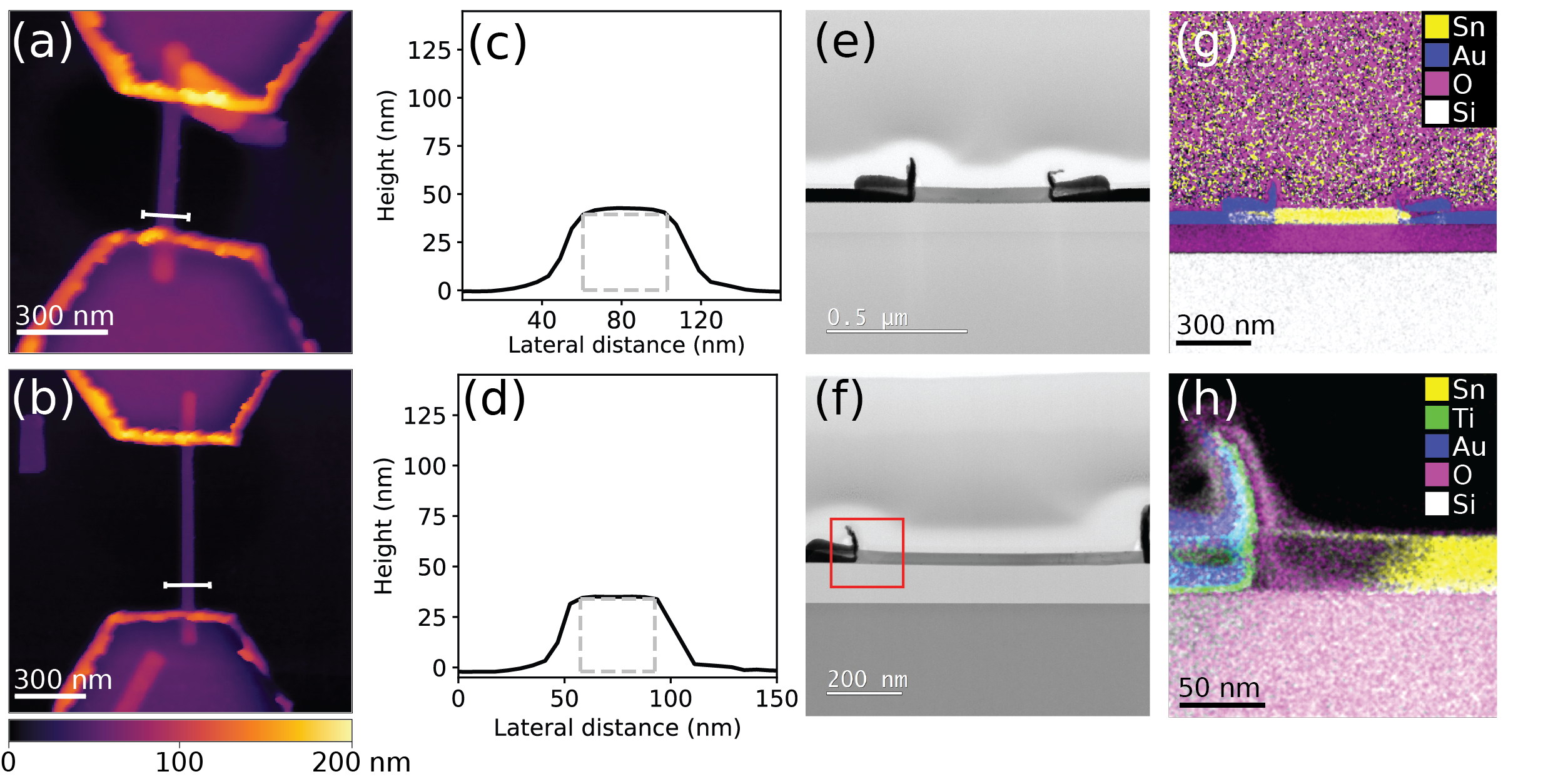}
\caption{\label{fig:morph}Two SnTe nanowire devices. \textbf{(a,b)} Atomic force microscopy (AFM) topography images of device C2 and A1 respectively. \textbf{(c,d)} Height profile measured across the nanowire as indicated by the white bar in (a) and (b) respectively. In grey we show the projected nanowire shape, which is square. \textbf{(e,f)} Bright-field scanning transmission electron microscopy micrographs of devices C1 and A1 respectively, cut along the long axis of the wire. \textbf{(g,h)} Energy-dispersive X-ray spectroscopy mappings of the wires shown in (e) and (f) respectively. For clearer visibility, Te is excluded from the overlay mapping, but is shown in the Supplementary Information as well as the separate mappings of the other elements. Panel (h) is zoomed-in on the red box shown in (f) and shows the presence of a Sn-poor pocket at the contact.}
\end{figure}

The morphology and composition of two SnTe nanowire devices are shown in Figure \ref{fig:morph}. Figures \ref{fig:morph}a,b show atomic force microscopfy (AFM) topography images. The Ti/Au contacts are visible in the top and bottom of the images, and the nanowire is situated in between the two contacts. The wires are straight and have a constant diameter of 40 nm $\pm$ 2 nm for Figure \ref{fig:morph}a and 35 nm $\pm$ 2 nm for Figure \ref{fig:morph}b.SnTe is known to easily oxidize on the surface \cite{li_surface_2017,neudachina_xps_2005,berchenko_surface_2018}, and therefore a sputter etch is performed to remove the self-limiting 3 nm native oxide on these nanowires, before depositing metal contacts. The deposited Ti/Au contacts are uniform, except for the edges of the contacts where the effects of side wall deposition during the sputtering fabrication process are visible. Figures \ref{fig:morph}c,d show the corresponding height profiles extracted from the AFM topography images in figures 1a,b. The height profile shows that the nanowire has a square cross-section, although it is a bit broadened in the lateral direction due to the finite sharpness of the AFM tip, resulting in a convolution between tip and sample. The projected nanowire shape is drawn in Figure \ref{fig:morph}c,d in the dashed grey line, defined by the flat top section of the AFM height profile, designated as the nanowire’s top facet. 

Figures \ref{fig:morph}e,f show bright-field scanning transmission electron microscopy (TEM) micrographs of two nanowire devices imaged orthogonal to the long axis of each wire. The TEM lamellas are slightly thicker than the nanowires themselves. Thus, the entire width of the nanowires is contained within the TEM samples, thereby avoiding structural damage to the nanowire by the sample preparation. As a result of the chosen TEM sample thickness, the metal contacts are visible both on top as well as on the sidewalls and in front of the nanowire end segments. A TEM close-up image of the contact area is shown in the Supplementary Information. Figure \ref{fig:morph}e shows a clean SnTe nanowire channel, exhibiting a pristine interior. The wire depicted in Figure \ref{fig:morph}f has a contrast difference at the left contact of the nanowire device, which suggests that the wire is thinner, contains less material, or has a different composition. Additional HR-TEM images of this region are presented in the Supplementary Information. Upon further investigating this, Figures \ref{fig:morph}g,h show the overlay of energy-dispersive X-ray spectroscopy (EDX) element mappings for the elements Sn, Au, O, and Si. The mappings of the separate elements can be found in the Supplementary Information, as well as the mapping of Te. Figure \ref{fig:morph}g shows a uniform Sn concentration along the entire wire, as well as underneath the contact. Figure \ref{fig:morph}h shows a non-uniform Sn concentration at the left contact, with a Sn-poor region directly next to the contact. It appears that Sn has diffused out of the wire, leaving a lower density, Sn-poor segment on this side of the nanowire. This is confirmed by an EDX compositional profile shown in the Supplementary Information showing that the Te:Sn ratio is 95:5 directly next to the contact. Because of a limited measurement resolution, we cannot conclude whether this segment is a binary Sn\textsubscript{x}Te\textsubscript{y} compound, or that the Sn counts originate from the oxidized outer shell of the nanowire. Furthermore, we observe that the Sn is mostly localized on the top and bottom of the cross-section where native oxide is formed. Elsewhere in the wire, the ratio of Sn to Te atoms is nearly equal, as expected from the growth stoichiometry. The strongly modified Sn to Te ratio below the contacts is likely formed during the device fabrication process, as no damage was observed in over fifteen nanowires just after growth. The two most invasive processes during device fabrication are PMMA-baking before e-beam lithography and the sputtering process, both of which cause an increase in the wire's temperature, potentially leading to solid-state diffusion. Sputtering, in particular, involves high power and directly affects the area where the contacts are deposited, as these regions are exposed during the process. Consequently, it is likely that the Sn-deficient segment observed in the device forms during the sputtering of Ti/Au contacts on the SnTe nanowire. 

We conclude from the results in Figure \ref{fig:morph} that we can fabricate reproducible nanowire devices in terms of their shape and size by inspecting their exterior. TEM inspection of the nanowire devices reveals variations in material composition between devices, probably introduced during the sputtering process. This variation can lead to distinct behavior in electronic transport, which we will examine in more detail in the remainder of this article. 

\subsection{Electronic classification} \label{sec:behav}

\begin{figure}[!ht]
\centering
\includegraphics[width=\linewidth]{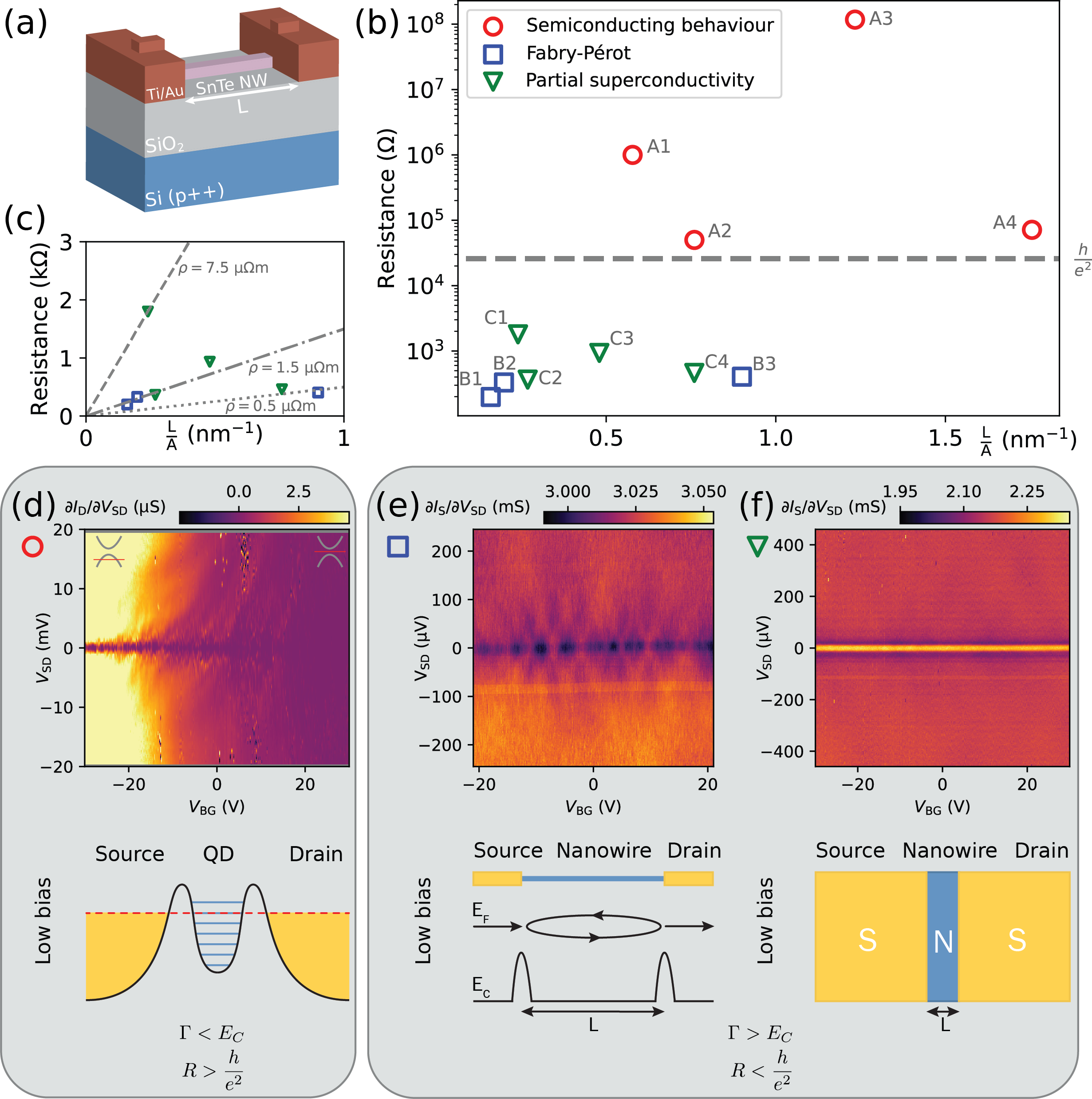}
\caption{\label{fig:transport}Electronic characterization of 11 nanowire devices. \textbf{(a)} Device schematic, with in red the source and drain contacts, in pink the nanowire on a Si/SiO\textsubscript{2} substrate. \textbf{(b)} The two-point resistance of the devices as function of the dimensional properties of the nanowires. The symbols indicate the different types of transport behavior. \textbf{(c)} Resistance as function of the dimensional properties in the low-resistance regime, with predicted linear relations for resistivity. \textbf{(d)} Conductance of device A1  showing clear depletion of the device above $V_\mathrm{{G}} = 20 \mathrm{V}$, with Coulomb blockaded regions at $V_\mathrm{{SD}} = \pm 2 \mathrm{mV}$. Schematic illustrates Coulomb blockade at low bias.  \textbf{(e)} Conductance of device B1 showing Fabry-Pérot resonances. The schematic shows Fabry-Pérot interference in a cavity with barriers at the contact interface. \textbf{(f)} Conductance of device C4 showing an onset of superconductivity around $V_\mathrm{{SD}} = 0 \mathrm{V}$. Schematic shows a hybrid superconductor - nanowire structure.}
\end{figure}

The schematic of a typical device for electronic transport measurements is displayed in Figure \ref{fig:transport}a. Measurements are carried out in a dilution refrigerator with a base temperature below 20 mK. Two-point measurements are carried out, either by applying a dc current or voltage bias to the source and drain contacts, depending on the resistance and noise level of the measurements. Furthermore, a voltage to the global Si/SiO\textsubscript{2} back gate is applied to study the field effect. Figure \ref{fig:transport}b shows the two-point resistances of the measured devices plotted against $\frac{L}{A}$, where $L$ is the channel length, and $A$ is the cross-sectional area of the wire. 
The resistance values are spread over more than five orders of magnitude, where we measure eight devices with a resistance $R < \frac{h}{e^2}$ and four devices with a resistance $R > \frac{h}{e^2}$. Additionally, Figure \ref{fig:transport}c shows the low-resistance regime plotted on a linear scale, showing there is no linear relation between the measured resistance and $\frac{L}{A}$. For metals and semiconductors a linear relation should be present, and therefore we relate the absence of a linear trend to the contact resistance of the devices. The colored markers in Figure \ref{fig:transport}b,c indicate three types of transport behavior in our measurements: semiconducting behavior, with in some cases the formation of an unintentional quantum dot; transport with the onset of superconductivity; and Fabry-Pérot oscillations. 

Figures \ref{fig:transport}d,e,f illustrate the three types of transport observed in the devices. IV curves for each device can be found in the Supplementary Information. Figure \ref{fig:transport}d presents an example of semiconducting behavior \cite{BeenakkerVanHouten}, measured in device A1. In this device, the differential conductance can be tuned from 9 \textmu S at $V_\mathrm{{G}} = -30$ V down to 0 at $V_\mathrm{{G}} = 30$ V. We observe similar semiconducting behavior in three other devices, labeled A2 to A4, and all have a resistance $R > \frac{h}{e^2}$ at $V_\mathrm{{G}} = 0$ V. Additionally, two of these devices show Coulomb diamonds, indicating transport through a quantum dot \cite{Hanson2007}. The graphic in Figure \ref{fig:transport}d shows the schematic band diagram of a quantum dot formed in between two barriers. In this regime it must hold that the coupling to the source and drain ($\Gamma$) is smaller than the charging energy ($E_\mathrm{C}$). Figure \ref{fig:QD} will explore the formation of the quantum dot in more detail.  

Figure \ref{fig:transport}e presents the transport measurement of device B1, displaying conductance oscillations that we attribute to Fabry-Pérot resonances. We observe similar oscillations in two other devices, indicated as B2 and B3 in Figure \ref{fig:transport}b. The illustrated schematic band diagram shows carrier interference in a nanowire device leading to Fabry-Pérot oscillations. In this interference regime it holds that $k_BT < E_C < \Gamma$ \cite{althuon_nano-assembled_2024}. Additionally, the mean free path ($l_e$) and the phase coherence length ($l_\phi$) must be larger than the channel length, indicating ballistic transport. To establish Fabry-Pérot oscillations a finite interface barrier is necessary, low enough to allow carriers to be injected into the channel but strong enough to act as back scatterers. The consecutive interface reflections allow for Fabry-Pérot resonances if the above conditions are met. From Hall effect measurements shown elsewhere \cite{manuscriptprep} we estimate a bulk mean free path of $l_{e}\approx$ 10 nm. Given this small length scale, 30 times smaller than the channel length, it is unexpected that such resonances occur in this system, which we will explore further in Figure \ref{fig:FP}.  

An example of the onset of superconductivity is depicted in Figure \ref{fig:transport}f, measured in device C4. We observe an enhancement in conductivity near zero bias voltage. The conductance remains constant across the entire gate voltage range. In three other devices we observe a conductance increase around zero bias as well, indicated in Figure \ref{fig:transport}b as devices C1 to C3. Above the critical field, Fabry-Pérot resonances are also visible in device C2. This onset of superconductivity will be examined in further detail in Figure \ref{fig:SC}.

The results in Figure \ref{fig:transport} show three different transport regimes, which we attribute to microscopic differences in transport at the metal-nanowire interface. Careful in-situ metal deposition after etching is performed, and from HR-TEM images (see Supplementary Information) we can conclude that the nanowire native oxide is removed. However, we cannot exclude the formation of TiO\textsubscript{2} at the contact interface as found in the EDX mapping for one of the devices (Supplementary Information). A varying amount of TiO\textsubscript{2} in the contact area can partially explain the variation in transport behavior, especially for the two devices in the high-resistance regime that do not show signs of the formation of a quantum dot. We observe no clear relation between the type of transport and the dimensional properties of the wire. We will now focus on the possible explanations for the establishment of the three transport mechanisms in these devices. 

\subsection{Hybrid superconductor - nanowire device} \label{sec:SC}
\begin{figure}[!ht]
\centering
\includegraphics[width=\linewidth]{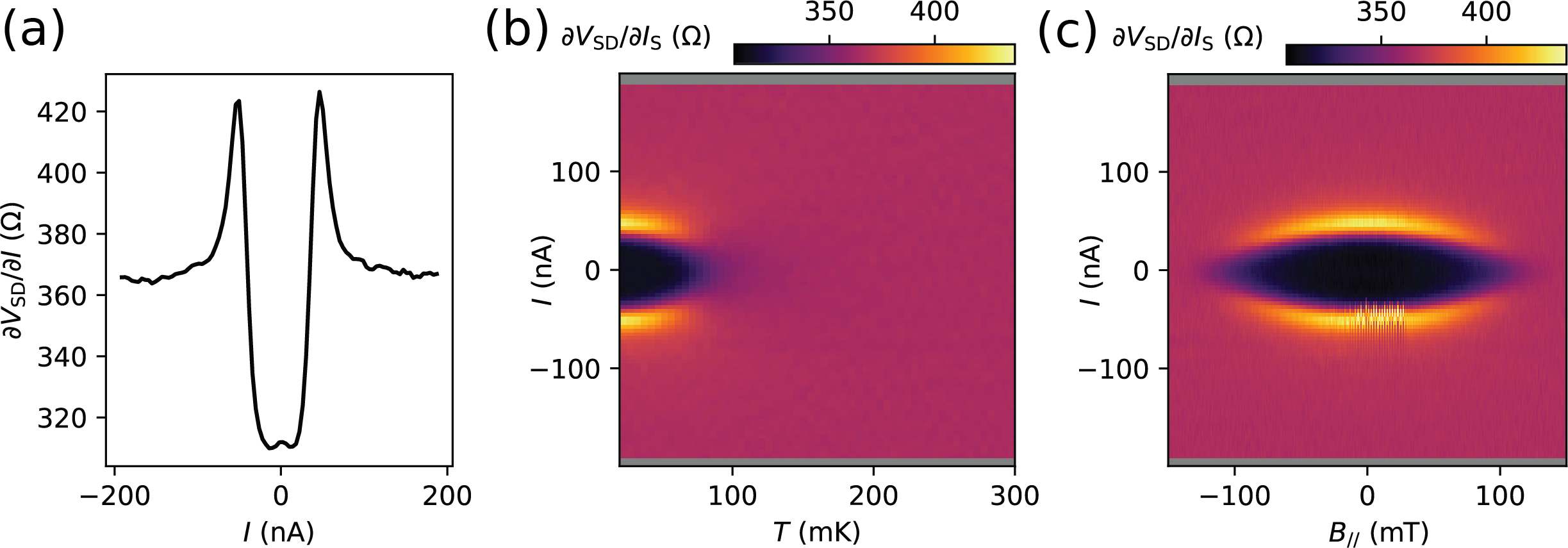}
\caption{\label{fig:SC} Partial superconductivity measured in a hybrid superconductor - nanowire device, specifically device C2. \textbf{(a)} Reduction of the differential resistance around zero bias at 50 mK. \textbf{(b)} Differential resistance as function of bias current and temperature, showing a critical temperature of 150 $\pm$ 10} mK. \textbf{(c)} Differential resistance as function of bias current and parallel magnetic field, showing a critical magnetic field of 140 $\pm$ 15 mT.
\end{figure}

Measurements of the critical field and critical temperature for device C2 are provided in Figure \ref{fig:SC}. Figure \ref{fig:SC}a shows the differential resistance as a function of the applied bias current. A region around zero applied bias shows a lowered resistance compared to the resistance at higher bias currents. The difference between the two is 55 $\Omega$. The critical temperature as depicted in Figure \ref{fig:SC}b is approximately 150 $\pm$ 10 mK.  Figure \ref{fig:SC}c indicates that the critical field is about 140 $\pm$ 15 mT. In this material system there are two main candidates for introducing superconductivity in the device: the SnTe nanowire and the 3 nm Ti sticking layer. The critical temperature of superconducting SnTe strongly depends on the carrier density, ranging from 30 mK to 210 mK as the carrier density increases \cite{Hein1969}. At our carrier density ($p_{3D} \approx 10^{20}$ cm\textsuperscript{-3}, for more information see Supplementary Information), the critical temperature of SnTe lies at the lower end of this range and does not align with the characteristics observed in our measurements. For Ti, the reported critical temperature of 0.39 K \cite{Titanium} also does not match the critical temperature determined in these measurements. However, the Ti sticking layer is thin, has a thick Au layer on top and may not be fully continuous, potentially introducing impurity scattering that could lower the critical temperature. To investigate further, we analyze the resistance drop in greater detail to determine whether it corresponds to Ti transitioning from a normal metal to a superconductor. Such a transition would reduce the total contact resistance between the Au lead and the wire. Considering the 3 nm thick Ti layer and the contact area between Ti and the nanowire, we calculate a Ti resistivity of $\rho \approx 4$ \textmu$\Omega \mathrm{cm}$. This value aligns with the reported resistivity of Ti at low temperatures \cite{Steele1953}, so we attribute the observed onset of superconductivity to the 3 nm Ti sticking layer for the contacts becoming superconducting and hence lowering the lead resistance, rather than the SnTe nanowire becoming superconducting either itself or because of the proximity effect. Josephson coupling can still be present below about a few nA considering the fact that the supercurrent can be smeared out by thermal energy in the system \cite{ambegaokar_voltage_1969} and would, therefore, not be detected. 

Only a few of the devices exhibit this partial superconducting behavior, suspectingly because of the variability in the thickness and coverage of the Ti layer in the nanowire devices. We thus conclude that the onset of superconductivity is an effect caused by a hybrid superconductor-nanowire device, where the Ti sticking layer causes a lowering in lead resistance. Next, we explore transport through a quantum dot.

\subsection{Te quantum dot} \label{sec:QD}
\begin{figure}[!ht]
\centering
\includegraphics[width=0.8\linewidth]{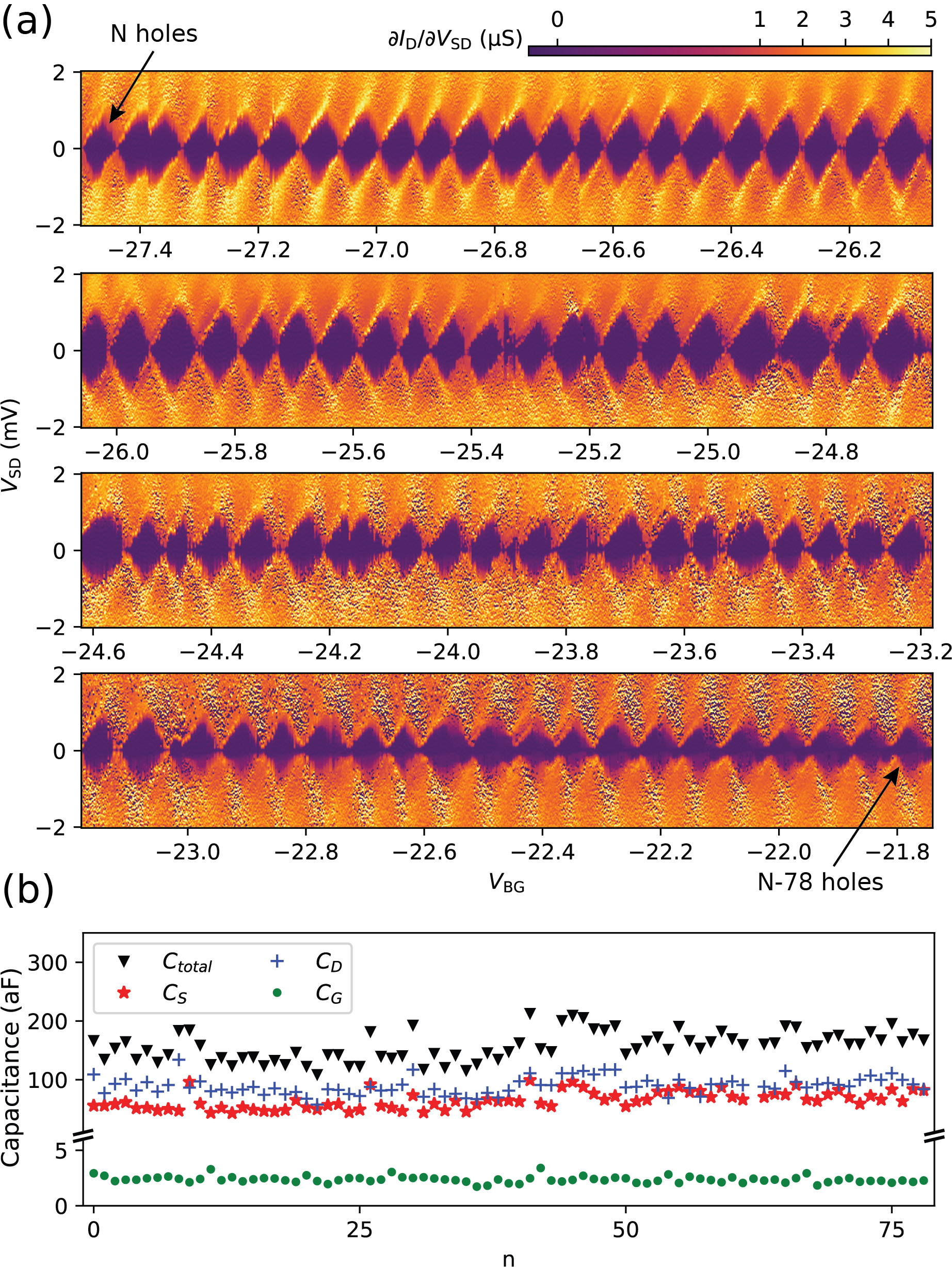}
\caption{\label{fig:QD} Measurement of a quantum dot in device A1. \textbf{(a)} Differential conductance as function of $V_\mathrm{{SD}}$ and $V_\mathrm{{BG}}$. This shows a stable measurement of 78 Coulomb blockaded diamonds. \textbf{(b)} Capacitance as function of the occupation number. All capacitances are extracted from the sizes of the diamonds shown in (a).}
\end{figure}

Figure \ref{fig:QD}a shows the differential conductance as function of backgate and source drain voltage, highlighting 78 distinct Coulomb blockaded diamonds in a SnTe nanowire device containing a quantum dot. The dot is particularly stable, as this measurement took roughly 15 hours, and there are only few charge jumps and instabilities in the map. To achieve Coulomb blockade in a device, certain requirements have to be met concerning the charging energy ($E_\mathrm{c}$), the conductance, and the tunnel coupling ($\Gamma$) to the source and drain contacts \cite{Beenakker1991}. The charging energy must significantly exceed the thermal energy, which it does in this case: the thermal energy is approximately 1.5 \textmu eV, while the charging energy is approximately three orders of magnitude higher. 
From the measured height of the Coulomb diamonds, we determine the charging energy to be 1.10 $\pm$ 0.15 meV. Estimating the size of the quantum dot requires knowledge of the gate capacitance, $C_\mathrm{G}$, which can be extracted from the width of the diamonds: $C_\mathrm{G} = \frac{e}{\Delta V_\mathrm{G}}$. The diamond width is characterized as 68 $\pm$ 8 mV, providing the basis for $C_\mathrm{G}$, as shown in Figure \ref{fig:QD}b as function of the dot occupation number, $N$. Furthermore, the negative and positive slopes of the diamonds determine the capacitance to the source $C_\mathrm{S}$ and to the drain $C_\mathrm{D}$ \cite{Hanson2007}, respectively. Next to the gate capacitance, $C_\mathrm{G}$, Figure \ref{fig:QD}b shows the source and drain capacitance, $C_\mathrm{S}$ and $C_\mathrm{D}$, as function of the dot occupation number, $N$, as well. From the gate capacitance, the length of the dot can be estimated with $L = \frac{C_\mathrm{G}t_\mathrm{ox}}{4\epsilon_0\epsilon_r r}$ \cite{Wunnicke2006}. A rough calculation tells that this corresponds to a dot length of 105 $\pm$ 15 nm. Since the channel length is 795 nm, the dot length is about eight times smaller, indicating that the quantum dot occupies only a small region between the two contacts. The size of the Sn-deficient region in Figure \ref{fig:morph}h is examined in more detail in a EDX compositional profile in the Supplementary Information, showing that at the left contact there is a small SnTe region underneath the contact, as well as at the right side of the compositional profile. The region in between, with a deficit of Sn, has a length of 110 $\pm$ 20 nm. Hence, we attribute the transport shown in Figure \ref{fig:QD} to the Te island at the contact. 

We conclude that a Sn-deficient part within the nanowire device can lead to the formation of a stable quantum dot, which behaves consistently with the constant-interaction model, as suggested by the extracted capacitances. Now we will focus on the establishment of Fabry-Pérot oscillations in different SnTe nanowire devices.

\subsection{Fabry-Pérot oscillations} \label{sec:FP}
\begin{figure}[!ht]
\centering
\includegraphics[width=0.8\linewidth]{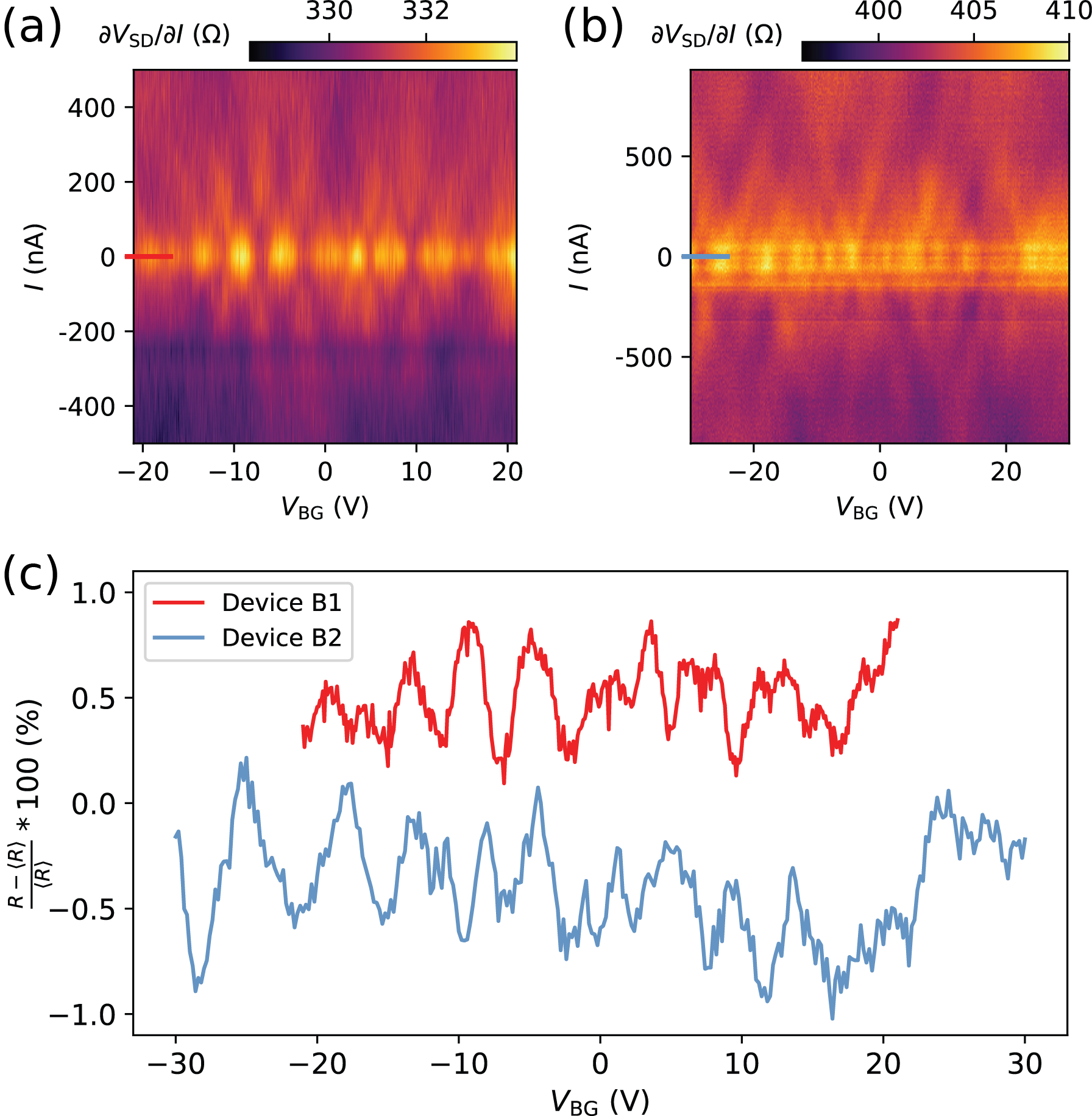}
\caption{\label{fig:FP} Measurement of Fabry-Pérot resonances in devices B1 and B2. \textbf{(a)} Differential resistance in device B1 as function as function of $I$ and $V_\mathrm{{BG}}$. \textbf{(b)} Differential resistance in device B2 as function as function of $I$ and $V_\mathrm{{BG}}$. \textbf{(c)} Line plot of the percentual difference in differential resistance at $I = 0$ as function of back gate voltage for the two devices. The two curves are offset from 0 for visibility.}
\end{figure}

Figure \ref{fig:FP}a,b show the numerical differential resistance as function of $I$ and $V_\mathrm{BG}$ of devices B1 and B2 respectively. The devices consist of nanowires from the same growth batch, but they are fabricated in different runs and measured in different cooldowns. Resonances manifest as a function of source-drain voltage and back-gate voltage, leading to an oscillation pattern with faint diamonds with a much lower resistance than typically observed for devices in the quantum dot regime. In both Figure \ref{fig:FP}a and Figure \ref{fig:FP}b we observe about ten oscillations in a back gate voltage range from -21 V to 21 V and -30 V to 30 V respectively. The diamond-like features are characteristic for Fabry-Pérot oscillations, where the $V_\mathrm{SD}$ and $V_\mathrm{BG}$ spacing between the diagonal lines is dependent on the geometry of the nanowire device \cite{Kretinin2010,liang_fabry_2001}. Device B1 and B2 have channel lengths of $\sim$350 nm. The previously estimated bulk mean free path of $\sim$10 nm is significantly smaller than the channel length of these devices, making it unlikely that Fabry-Pérot resonances establish in the bulk channel of the material. Nevertheless, we qualitatively see similar results for the four devices (B1 to B3, and C2) as indicated in Figure \ref{fig:transport}b. Therefore, we believe that the Fabry-Pérot oscillations here are an effect of the 2D surface states parallel to the high bulk conductivity. The interference pattern can be modeled using the Fermi velocity and quantization along the length and width directions, where the junction's physical dimensions, extracted from AFM images, are used. The energy levels are described by \cite{oksanen_single_2014}: 
\begin{equation}
    E_{n_L,n_w} = \pm \frac{\hbar v_F}{2L}\sqrt{n_L^2 + n_w^2\left(\frac{L}{4w}\right)^2}, \label{eq:FP}
\end{equation}
with $n_L$ and $n_w$ representing the quantized subbands along the length ($L$) and width ($w$) directions, respectively. Using the model, the Fabry-Pérot oscillations can be reproduced, and energy scales comparable to the experimental results are extracted. Detailed modeling of these oscillations, based on the parameters of the fabricated and measured devices, is presented in the Supplementary Information.  
Figure \ref{fig:FP}c shows the line traces of the normalized resistance as a function of gate voltage at $I = 0$, where $\langle R\rangle$ is the mean resistance of the entire trace. Both devices exhibit resistance changes of 1.0 $\pm$ 0.2 \% of the total resistance, induced by the resonances. Such a small percentage is consistent with the hypothesis that the Fabry-Pérot oscillations arise from a two-dimensional transport contribution superimposed on a dominant three-dimensional bulk conductance. The high bulk conductance thus reduces the relative impact of the resonances on the total device resistance. We interpret this system as consisting of two parallel resistive paths: a 3D bulk resistor and a 2D Fabry-Pérot resistor, from which we extract a parallel Fabry-Pérot resistor value of 47 $\pm$ 10 k$\Omega$ corresponding to $\sim0.55$ $\mathrm{e^2/h}$. 

From Figure \ref{fig:FP}, we conclude that ballistic transport is present in the devices exhibiting Fabry-Pérot oscillations. Considering surface state transport, the energy scales of the experimental data and model correspond well for our channel length and width. Therefore, we attribute the observed Fabry-Pérot oscillations to ballistic transport occurring on the surface of the SnTe nanowires.

\section{Conclusion}\label{conclusion}
We have demonstrated the fabrication of SnTe nanowire devices, exhibiting three distinct types of quantum transport behavior: semiconducting characteristics, Fabry-Pérot resonances, and the onset superconductivity. Each of these phenomena are attributed to physical effects at the metal-nanowire contact. Specifically, the formation of a highly stable quantum dot is explained by the presence of a Te island at the contact interface. Additionally, partial superconductivity is observed, attributed to the presence of the Ti sticking layer, rendering a hybrid superconductor-nanowire device when the Ti becomes superconducting. For finite contact transparencies, multiple devices exhibit Fabry-Pérot resonances, arising from weak reflections of forward-propagating modes at the contacts, which provides a signature of ballistic transport in SnTe nanowire surface states. 

This diverse spectrum of electronic transport behaviors indicates variability in the sputtering process, highlighting the need to refine the etching procedure for SnTe nanowires and to possibly explore the necessity of introducing post-annealing after device fabrication. Nevertheless, our results demonstrate the potential of SnTe nanowires as a promising platform for exploring Josephson physics, as even a thin layer of superconducting material can partially induce superconductivity. Moreover, ballistic transport via the 2D surface states paves the way for a variety of follow-up experiments and novel device architectures based on SnTe nanowires, including exploring topological superconductivity and the development of low-energy spintronic devices.

\section{Experimental section} \label{methods}
\textsl{Self-assisted nanowire growth \cite{Mientjes2024}: } SnTe nanowires are grown on a [100]-oriented Si substrate with a 20 nm SiN\textsubscript{x} mask. In the mask, holes are patterned and etched using reactive ion etching. Growth happens selectively out-of-plane in an ultra-high vacuum (3-6 $\times 10^{-10}$ mbar) MBE system. Material is deposited from a Sn effusion cell and a Te cracker cell under a 30-degree incident angle with respect to the substrate normal. The flux ratio of group IV to group VI is 0.16. The growth recipe consists of a 2 min Sn pre-deposition followed by 220 min of SnTe deposition. During the entire growth recipe, the substrate is kept at a temperature of 339 \degree C and is rotating at 5.3 RPM. 

\textsl{Device fabrication: } The wires are deposited on a p\textsuperscript{++} doped Si substrate covered with 100 nm SiO\textsubscript{2} and contacted after SEM imaging. Source and drain contacts are defined using electron beam lithography. After development, an in-situ sputter etch is performed to remove the native oxide and 3 nm / 50 nm Ti/Au contacts are sputtered. Lift-off is performed in a dimethylsulfoxide bath heated to \SI{60}{\celsius}.

\textsl{Cryogenic measurements: } Measurements are carried out in a dilution refrigerator with a base temperature below 20 mK, equipped with a 6-1-1 T vector magnet. Measurements are carried out at base temperature, unless stated otherwise. Using an IVVI-DAC2-rack with PC-controlled DACs and amplification modules, both DC current and voltage bias are applied to the source and drain contacts, depending on the resistance and noise level of the measurements. Furthermore, a voltage to the global Si/SiO\textsubscript{2} back gate is applied using a Keithley2400 to study the field effect on the devices.

\textsl{TEM sample preparation and studies: } After performing transport measurements, TEM lamellas were made using a Thermo Fisher Helios 5C Dual Beam. Prior to FIB milling, protective carbon layers were deposited by electron and ion beam deposition, respectively. Subsequent TEM studies were performed using a JEOL ARM 200F Transmission Electron Microscope, probe corrected, equipped with a 100 mm2 Centurio SDD EDX detector, operated at 200 kV. EDX elemental mappings of 256x256 resolution were acquired and subsequently quantified per pixel after applying 3x3 binning using the JEOL Analysis Station software. The quantification procedure has been discussed in the Supplementary Data of Ref. \cite{Mientjes2024}. 

\section{Acknowledgements}
This work has been supported from the project "HOTNANO" by the Dutch Organization for Scientific Research (OCENW.GROOT.2019.004), the European Research Council (ERC TOCINA 834290), Solliance and the Dutch province of Noord-Brabant for funding the TEM facility.

\section{Data availability}
The data that support the findings of this study are openly available in Zenodo at http://doi.org/10.5281/zenodo.14959524., reference number \cite{witmans_2025_14959524}.
 
\newpage

\bibliography{references,SnTegeneralwork_v2}

\begin{thebibliography}{49}
\providecommand{\natexlab}[1]{#1}
\providecommand{\url}[1]{\texttt{#1}}
\expandafter\ifx\csname urlstyle\endcsname\relax
  \providecommand{\doi}[1]{doi: #1}\else
  \providecommand{\doi}{doi: \begingroup \urlstyle{rm}\Url}\fi

\bibitem[Kane and Mele(2005)]{Kane2005}
Charles~L Kane and Eugene~J Mele.
\newblock Z 2 topological order and the quantum spin hall effect.
\newblock \emph{Physical review letters}, 95\penalty0 (14):\penalty0 146802, 2005.
\newblock \doi{10.1103/PhysRevLett.95.146802}.

\bibitem[Qi and Zhang(2011)]{ReviewTI2011}
Xiao-Liang Qi and Shou-Cheng Zhang.
\newblock Topological insulators and superconductors.
\newblock \emph{Rev. Mod. Phys.}, 83:\penalty0 1057--1110, 10 2011.
\newblock \doi{10.1103/RevModPhys.83.1057}.
\newblock URL \url{https://link.aps.org/doi/10.1103/RevModPhys.83.1057}.

\bibitem[Fu et~al.(2007)Fu, Kane, and Mele]{FuKaneMele2007}
Liang Fu, C.~L. Kane, and E.~J. Mele.
\newblock Topological insulators in three dimensions.
\newblock \emph{Phys. Rev. Lett.}, 98:\penalty0 106803, 03 2007.
\newblock \doi{10.1103/PhysRevLett.98.106803}.
\newblock URL \url{https://link.aps.org/doi/10.1103/PhysRevLett.98.106803}.

\bibitem[Fu and Kane(2007)]{Fu2007}
Liang Fu and C.~L. Kane.
\newblock Topological insulators with inversion symmetry.
\newblock \emph{Phys. Rev. B}, 76:\penalty0 045302, 07 2007.
\newblock \doi{10.1103/PhysRevB.76.045302}.
\newblock URL \url{https://link.aps.org/doi/10.1103/PhysRevB.76.045302}.

\bibitem[König et~al.(2007)König, Wiedmann, Brüne, Roth, Buhmann, Molenkamp, Qi, and Zhang]{Konig2007}
Markus König, Steffen Wiedmann, Christoph Brüne, Andreas Roth, Hartmut Buhmann, Laurens~W. Molenkamp, Xiao-Liang Qi, and Shou-Cheng Zhang.
\newblock Quantum spin hall insulator state in {H}g{T}e quantum wells.
\newblock \emph{Science}, 318\penalty0 (5851):\penalty0 766--70, 11 2007.
\newblock \doi{10.1126/science.1148047}.

\bibitem[Moore(2010)]{Moore2010}
Joel~E. Moore.
\newblock The birth of topological insulators.
\newblock \emph{Nature}, 464\penalty0 (7268):\penalty0 194--198, 03 2010.
\newblock \doi{10.1038/nature08916}.

\bibitem[Fu(2011)]{Fu2011}
Liang Fu.
\newblock Topological crystalline insulators.
\newblock \emph{Phys. Rev. Lett.}, 106:\penalty0 106802, 03 2011.
\newblock \doi{10.1103/PhysRevLett.106.106802}.
\newblock URL \url{https://link.aps.org/doi/10.1103/PhysRevLett.106.106802}.

\bibitem[Hsieh et~al.(2012)Hsieh, Lin, Lin, Duan, Bansil, and Fu]{Hseih2012}
Timothy~H. Hsieh, Hsin Lin, Junwei Lin, Wenhui Duan, Arun Bansil, and Liang Fu.
\newblock Topological crystalline insulators in the {SnTe} material class.
\newblock \emph{Nature Communications}, 3\penalty0 (982), 07 2012.
\newblock \doi{10.1038/ncomms1969}.

\bibitem[Pankratov et~al.(1987)Pankratov, Pakhomov, and Volkov]{PANKRATOV198793}
O.A. Pankratov, S.V. Pakhomov, and B.A. Volkov.
\newblock Supersymmetry in heterojunctions: Band-inverting contact on the basis of pb1-xsnxte and hg1-xcdxte.
\newblock \emph{Solid State Communications}, 61\penalty0 (2):\penalty0 93--96, 1987.
\newblock ISSN 0038-1098.
\newblock \doi{https://doi.org/10.1016/0038-1098(87)90934-3}.
\newblock URL \url{https://www.sciencedirect.com/science/article/pii/0038109887909343}.

\bibitem[Bauer~Pereira et~al.(2013)Bauer~Pereira, Sergueev, Gorsse, Dadda, Müller, and Hermann]{Bauer2013}
Paula Bauer~Pereira, Ilya Sergueev, Stéphane Gorsse, Jayaram Dadda, Eckhard Müller, and Raphaël~P. Hermann.
\newblock Lattice dynamics and structure of {GeTe}, {SnTe} and {PbTe}.
\newblock \emph{physica status solidi (b)}, 250\penalty0 (7):\penalty0 1300--1307, 2013.
\newblock \doi{https://doi.org/10.1002/pssb.201248412}.
\newblock URL \url{https://onlinelibrary.wiley.com/doi/abs/10.1002/pssb.201248412}.

\bibitem[Wang et~al.(2014)Wang, West, Liu, Li, Yan, Gu, Zhang, and Duan]{Snvacancy2014}
Na~Wang, Damien West, Junwei Liu, Jia Li, Qimin Yan, Bing-Lin Gu, S.~B. Zhang, and Wenhui Duan.
\newblock Microscopic origin of the $p$-type conductivity of the topological crystalline insulator snte and the effect of pb alloying.
\newblock \emph{Phys. Rev. B}, 89:\penalty0 045142, 01 2014.
\newblock \doi{10.1103/PhysRevB.89.045142}.
\newblock URL \url{https://link.aps.org/doi/10.1103/PhysRevB.89.045142}.

\bibitem[Dybko et~al.(2017)Dybko, Szot, Szczerbakow, Gutowska, Zajarniuk, Domagala, Szewczyk, Story, and Zawadzki]{dybko_experimental_2017}
K.~Dybko, M.~Szot, A.~Szczerbakow, M.U. Gutowska, T.~Zajarniuk, J.Z. Domagala, A.~Szewczyk, T.~Story, and W.~Zawadzki.
\newblock Experimental evidence for topological surface states wrapping around a bulk {SnTe} crystal.
\newblock \emph{Physical Review B}, 96\penalty0 (20), 2017.
\newblock \doi{10.1103/PhysRevB.96.205129}.
\newblock URL \url{https://www.scopus.com/inward/record.uri?eid=2-s2.0-85039933076&doi=10.1103%2fPhysRevB.96.205129&partnerID=40&md5=c9bfdfa5d2df3e6fa8397e1afb8f10c5}.

\bibitem[Costa et~al.(2021)Costa, Mengui, Abramof, Rappl, Soares, De~Castro, and Peres]{costa_investigation_2021}
IF~Costa, UA~Mengui, E~Abramof, PHO Rappl, DAW Soares, S~De~Castro, and ML~Peres.
\newblock Investigation of {Shubnikov}-de {Haas} oscillations in a crystalline topological insulator {SnTe}/{Sn1}-{xEuxTe} heterostructure.
\newblock \emph{PHYSICAL REVIEW B}, 104\penalty0 (12), September 2021.
\newblock ISSN 2469-9950.
\newblock \doi{10.1103/PhysRevB.104.125203}.

\bibitem[Okazaki et~al.(2018)Okazaki, Wiedmann, Pezzini, Peres, Rappl, and Abramof]{okazaki_shubnikov-haas_2018}
A.K. Okazaki, S.~Wiedmann, S.~Pezzini, M.L. Peres, P.H.O. Rappl, and E.~Abramof.
\newblock Shubnikov-de {Haas} oscillations in topological crystalline insulator {SnTe}(111) epitaxial films.
\newblock \emph{Physical Review B}, 98\penalty0 (19), 2018.
\newblock \doi{10.1103/PhysRevB.98.195136}.
\newblock URL \url{https://www.scopus.com/inward/record.uri?eid=2-s2.0-85057417006&doi=10.1103%2fPhysRevB.98.195136&partnerID=40&md5=b9171adee9e0fa1b75a1924d5129f468}.

\bibitem[Taskin et~al.(2014)Taskin, Yang, Sasaki, Segawa, and Ando]{taskin_topological_2014}
A.A. Taskin, F.~Yang, S.~Sasaki, K.~Segawa, and Y.~Ando.
\newblock Topological surface transport in epitaxial {SnTe} thin films grown on {Bi} 2 {Te} 3.
\newblock \emph{Physical Review B - Condensed Matter and Materials Physics}, 89\penalty0 (12), 2014.
\newblock \doi{10.1103/PhysRevB.89.121302}.
\newblock URL \url{https://www.scopus.com/inward/record.uri?eid=2-s2.0-84896927619&doi=10.1103%2fPhysRevB.89.121302&partnerID=40&md5=2205c15fd2b86a84a3d5a71b90bf0bdf}.

\bibitem[Safdar et~al.(2013)Safdar, Wang, Mirza, Wang, Xu, and He]{Safdar2013}
Muhammad Safdar, Qisheng Wang, Misbah Mirza, Zhenxing Wang, Kai Xu, and Jun He.
\newblock Topological surface transport properties of single-crystalline snte nanowire.
\newblock \emph{Nano Letters}, 13\penalty0 (11):\penalty0 5344--5349, 11 2013.
\newblock \doi{10.1021/nl402841x}.

\bibitem[Liu et~al.(2024)Liu, Zhang, Li, Zhang, Liu, Narayan, Chen, Zou, and Xiu]{liu_cr_2024}
S.~Liu, E.~Zhang, Z.~Li, X.~Zhang, W.~Liu, A.~Narayan, Z.-G. Chen, J.~Zou, and F.~Xiu.
\newblock Cr doping-induced ferromagnetism in {SnTe} thin films.
\newblock \emph{npj Quantum Materials}, 9\penalty0 (1), 2024.
\newblock \doi{10.1038/s41535-024-00667-x}.
\newblock URL \url{https://www.scopus.com/inward/record.uri?eid=2-s2.0-85199147835&doi=10.1038%2fs41535-024-00667-x&partnerID=40&md5=bfcb49e73177b2ea8eaa9122c46be276}.

\bibitem[Assaf et~al.(2014)Assaf, Katmis, Wei, Satpati, Zhang, Bennett, Harris, Moodera, and Heiman]{assaf_quantum_2014}
B.A. Assaf, F.~Katmis, P.~Wei, B.~Satpati, Z.~Zhang, S.P. Bennett, V.G. Harris, J.S. Moodera, and D.~Heiman.
\newblock Quantum coherent transport in {SnTe} topological crystalline insulator thin films.
\newblock \emph{Applied Physics Letters}, 105\penalty0 (10), 2014.
\newblock \doi{10.1063/1.4895456}.

\bibitem[Zou et~al.(2019)Zou, Albright, Dagdeviren, Morales-Acosta, Simon, Zhou, Mandal, Ismail-Beigi, Schwarz, Altman, Walker, and Ahn]{zou_revealing_2019}
K.~Zou, S.D. Albright, O.E. Dagdeviren, M.D. Morales-Acosta, G.H. Simon, C.~Zhou, S.~Mandal, S.~Ismail-Beigi, U.D. Schwarz, E.I. Altman, F.J. Walker, and C.H. Ahn.
\newblock Revealing surface-state transport in ultrathin topological crystalline insulator {SnTe} films.
\newblock \emph{APL Materials}, 7\penalty0 (5), 2019.
\newblock \doi{10.1063/1.5096279}.
\newblock URL \url{https://www.scopus.com/inward/record.uri?eid=2-s2.0-85065833256&doi=10.1063%2f1.5096279&partnerID=40&md5=71212712484715df4e47b4b03ecfa978}.

\bibitem[Yan et~al.(2020)Yan, Wei, Bai, Wang, Zhang, Ma, Liu, and Zhang]{yan_structure_2020}
C.H. Yan, F.~Wei, Y.~Bai, F.~Wang, A.Q. Zhang, S.~Ma, W.~Liu, and Z.D. Zhang.
\newblock Structure and topological transport in {Pb}-doping topological crystalline insulator {SnTe} (001) film.
\newblock \emph{Journal of Materials Science and Technology}, 44:\penalty0 223--228, 2020.
\newblock \doi{10.1016/j.jmst.2019.10.033}.

\bibitem[Akiyama et~al.(2016)Akiyama, Fujisawa, Yamaguchi, Ishikawa, and Kuroda]{akiyama_two-dimensional_2016}
R.~Akiyama, K.~Fujisawa, T.~Yamaguchi, R.~Ishikawa, and S.~Kuroda.
\newblock Two-dimensional quantum transport of multivalley (111) surface state in topological crystalline insulator {SnTe} thin films.
\newblock \emph{Nano Research}, 9\penalty0 (2):\penalty0 490--498, 2016.
\newblock \doi{10.1007/s12274-015-0930-8}.
\newblock URL \url{https://www.scopus.com/inward/record.uri?eid=2-s2.0-84958102060&doi=10.1007%2fs12274-015-0930-8&partnerID=40&md5=913c22c2e24af19b4e30b76c2ca5f4b8}.

\bibitem[De~Castro et~al.(2022)De~Castro, Kawata, Lopes, Rappl, Abramof, and Peres]{de_castro_weak_2022}
S.~De~Castro, B.~Kawata, G.R.F. Lopes, P.H.D.O. Rappl, E.~Abramof, and M.L. Peres.
\newblock Weak antilocalization effect and multi-channel transport in {SnTe} quantum well.
\newblock \emph{Applied Physics Letters}, 120\penalty0 (20), 2022.
\newblock \doi{10.1063/5.0088499}.

\bibitem[Akiyama et~al.(2014)Akiyama, Fujisawa, Sakurai, and Kuroda]{akiyama_weak_2014}
R.~Akiyama, K.~Fujisawa, R.~Sakurai, and S.~Kuroda.
\newblock Weak antilocalization in (111) thin films of a topological crystalline insulator {SnTe}.
\newblock volume 568, 2014.
\newblock \doi{10.1088/1742-6596/568/5/052001}.

\bibitem[Li et~al.(2022)Li, Yang, Wang, Zhu, Qu, Wang, and Yang]{li_weak_2022}
X.~Li, Y.~Yang, X.~Wang, P.~Zhu, F.~Qu, Z.~Wang, and F.~Yang.
\newblock Weak {Antilocalization} in {Polycrystalline} {SnTe} {Films} {Deposited} by {Magnetron} {Sputtering}.
\newblock \emph{Crystals}, 12\penalty0 (6), 2022.
\newblock \doi{10.3390/cryst12060773}.

\bibitem[Albright et~al.(2021)Albright, Zou, Walker, and Ahn]{albright_weak_2021}
S.D. Albright, K.~Zou, F.J. Walker, and C.H. Ahn.
\newblock Weak antilocalization in topological crystalline insulator {SnTe} films deposited using amorphous seeding on {SrTiO3}.
\newblock \emph{APL Materials}, 9\penalty0 (11), 2021.
\newblock \doi{10.1063/5.0065627}.

\bibitem[Khaliq et~al.(2022)Khaliq, Dziawa, Minikaev, Arciszewska, Avdonin, Brodowska, and Kilanski]{khaliq_low_2022}
A.~Khaliq, P.~Dziawa, R.~Minikaev, M.~Arciszewska, A.~Avdonin, B.~Brodowska, and L.~Kilanski.
\newblock Low {Temperature} {Weak} {Anti}-{Localization} {Effect} in the {GeTe} and {SnTe} {Epitaxial} {Layers}.
\newblock \emph{Acta Physica Polonica A}, 142\penalty0 (5):\penalty0 657--661, 2022.
\newblock \doi{10.12693/APhysPolA.142.657}.

\bibitem[Klett et~al.(2018)Klett, Schönle, Becker, Dyck, Borisov, Rott, Ramermann, Büker, Haskenhoff, Krieft, Hübner, Reimer, Shekhar, Schmalhorst, Hütten, Felser, Wernsdorfer, and Reiss]{klett_proximity-induced_2018}
R.~Klett, J.~Schönle, A.~Becker, D.~Dyck, K.~Borisov, K.~Rott, D.~Ramermann, B.~Büker, J.~Haskenhoff, J.~Krieft, T.~Hübner, O.~Reimer, C.~Shekhar, J.-M. Schmalhorst, A.~Hütten, C.~Felser, W.~Wernsdorfer, and G.~Reiss.
\newblock Proximity-{Induced} {Superconductivity} and {Quantum} {Interference} in {Topological} {Crystalline} {Insulator} {SnTe} {Thin}-{Film} {Devices}.
\newblock \emph{Nano Letters}, 18\penalty0 (2):\penalty0 1264--1268, 2018.
\newblock \doi{10.1021/acs.nanolett.7b04870}.
\newblock URL \url{https://www.scopus.com/inward/record.uri?eid=2-s2.0-85042100747&doi=10.1021%2facs.nanolett.7b04870&partnerID=40&md5=a13787ed40ee4c1afab07552a5c3dfed}.

\bibitem[Liu et~al.(2021)Liu, Han, Wei, Hynek, Hart, Han, Trimble, Williams, Zhu, and Cha]{liu_synthesis_2021}
P.~Liu, H.J. Han, J.~Wei, D.~Hynek, J.L. Hart, M.G. Han, C.J. Trimble, J.~Williams, Y.~Zhu, and J.J. Cha.
\newblock Synthesis of {Narrow} {SnTe} {Nanowires} {Using} {Alloy} {Nanoparticles}.
\newblock \emph{ACS Applied Electronic Materials}, 3\penalty0 (1):\penalty0 184--191, 2021.
\newblock \doi{10.1021/acsaelm.0c00740}.

\bibitem[Trimble et~al.(2021)Trimble, Wei, Yuan, Kalantre, Liu, Han, Han, Zhu, Cha, Fu, and Williams]{Trimble2021}
C.J. Trimble, M.T. Wei, N.F.Q. Yuan, S.S. Kalantre, P.~Liu, H.-J. Han, M.-G. Han, Y.~Zhu, J.J. Cha, L.~Fu, and J.R. Williams.
\newblock Josephson detection of time-reversal symmetry broken superconductivity in {SnTe} nanowires.
\newblock \emph{npj Quantum Materials}, 6\penalty0 (61), 5 2021.
\newblock \doi{10.1038/s41535-021-00359-w}.

\bibitem[Baral and Lakhani(2019)]{baral_growth_2019}
S.~Baral and A.~Lakhani.
\newblock Growth and characterization of snte crystalline topological insulator.
\newblock volume 2162, 2019.
\newblock \doi{10.1063/1.5130336}.
\newblock URL \url{https://www.scopus.com/inward/record.uri?eid=2-s2.0-85074767655&doi=10.1063%2f1.5130336&partnerID=40&md5=1f22deba74486515d569c0950da7b027}.

\bibitem[Shen et~al.(2014)Shen, Jung, Disa, Walker, Ahn, and Cha]{shen_synthesis_2014}
J.~Shen, Y.~Jung, A.S. Disa, F.J. Walker, C.H. Ahn, and J.J. Cha.
\newblock Synthesis of {SnTe} nanoplates with \{100\} and \{111\} surfaces.
\newblock \emph{Nano Letters}, 14\penalty0 (7):\penalty0 4183--4188, 2014.
\newblock \doi{10.1021/nl501953s}.
\newblock URL \url{https://www.scopus.com/inward/record.uri?eid=2-s2.0-84903957041&doi=10.1021%2fnl501953s&partnerID=40&md5=72679f65ef7ed42daf8611bc9d302eb6}.

\bibitem[Mientjes et~al.(2024)Mientjes, Guan, Lueb, Verheijen, and Bakkers]{Mientjes2024}
Mathijs G.~C. Mientjes, Xin Guan, Pim~J.H. Lueb, Marcel~A. Verheijen, and Erik P. A.~M. Bakkers.
\newblock Catalyst-free mbe growth of pbsnte nanowires with tunable aspect ratio.
\newblock \emph{Nanotechnology}, 35, 05 2024.
\newblock \doi{10.1088/1361-6528/ad47c8}.

\bibitem[Li et~al.(2017)Li, Xu, Losovyj, Li, Chen, Swartzentruber, Sinitsyn, Yoo, Jia, and Zhang]{li_surface_2017}
Zhen Li, Enzhi Xu, Yaroslav Losovyj, Nan Li, Aiping Chen, Brian Swartzentruber, Nikolai Sinitsyn, Jinkyoung Yoo, Quanxi Jia, and Shixiong Zhang.
\newblock Surface oxidation and thermoelectric properties of indium-doped tin telluride nanowires.
\newblock \emph{Nanoscale}, 9\penalty0 (35):\penalty0 13014--13024, September 2017.
\newblock ISSN 2040-3372.
\newblock \doi{10.1039/c7nr04934j}.

\bibitem[Neudachina et~al.(2005)Neudachina, Shatalova, Shtanov, Yashina, Zyubina, Tamm, and Kobeleva]{neudachina_xps_2005}
V.~S. Neudachina, T.~B. Shatalova, V.~I. Shtanov, L.~V. Yashina, T.~S. Zyubina, M.~E. Tamm, and S.~P. Kobeleva.
\newblock {XPS} study of {SnTe}(1   0   0) oxidation by molecular oxygen.
\newblock \emph{Surface Science}, 584\penalty0 (1):\penalty0 77--82, June 2005.
\newblock ISSN 0039-6028.
\newblock \doi{10.1016/j.susc.2005.01.061}.
\newblock URL \url{https://www.sciencedirect.com/science/article/pii/S0039602805003584}.

\bibitem[Berchenko et~al.(2018)Berchenko, Vitchev, Trzyna, Wojnarowska-Nowak, Szczerbakow, Badyla, Cebulski, and Story]{berchenko_surface_2018}
N.~Berchenko, R.~Vitchev, M.~Trzyna, R.~Wojnarowska-Nowak, A.~Szczerbakow, A.~Badyla, J.~Cebulski, and T.~Story.
\newblock Surface oxidation of {SnTe} topological crystalline insulator.
\newblock \emph{Applied Surface Science}, 452:\penalty0 134--140, September 2018.
\newblock ISSN 0169-4332.
\newblock \doi{10.1016/j.apsusc.2018.04.246}.
\newblock URL \url{https://www.sciencedirect.com/science/article/pii/S0169433218312261}.

\bibitem[Beenakker and {van Houten}(1991)]{BeenakkerVanHouten}
C.W.J. Beenakker and H.~{van Houten}.
\newblock Quantum transport in semiconductor nanostructures.
\newblock In Henry Ehrenreich and David Turnbull, editors, \emph{Semiconductor Heterostructures and Nanostructures}, volume~44 of \emph{Solid State Physics}, pages 1--228. Academic Press, 1991.
\newblock \doi{https://doi.org/10.1016/S0081-1947(08)60091-0}.
\newblock URL \url{https://www.sciencedirect.com/science/article/pii/S0081194708600910}.

\bibitem[Hanson et~al.(2007)Hanson, Kouwenhoven, Petta, Tarucha, and Vandersypen]{Hanson2007}
R.~Hanson, L.~P. Kouwenhoven, J.~R. Petta, S.~Tarucha, and L.~M.~K. Vandersypen.
\newblock Spins in few-electron quantum dots.
\newblock \emph{Rev. Mod. Phys.}, 79:\penalty0 1217--1265, 10 2007.
\newblock \doi{10.1103/RevModPhys.79.1217}.
\newblock URL \url{https://link.aps.org/doi/10.1103/RevModPhys.79.1217}.

\bibitem[Althuon et~al.(2024)Althuon, Cubaynes, Auer, Sürgers, and Wernsdorfer]{althuon_nano-assembled_2024}
Tim Althuon, Tino Cubaynes, Aljoscha Auer, Christoph Sürgers, and Wolfgang Wernsdorfer.
\newblock Nano-assembled open quantum dot nanotube devices.
\newblock \emph{Communications Materials}, 5\penalty0 (1):\penalty0 1--7, January 2024.
\newblock ISSN 2662-4443.
\newblock \doi{10.1038/s43246-023-00439-3}.
\newblock URL \url{https://www.nature.com/articles/s43246-023-00439-3}.
\newblock Publisher: Nature Publishing Group.

\bibitem[man()]{manuscriptprep}
Manuscript in preparation.

\bibitem[HEIN and MEIJER(1969)]{Hein1969}
R.~A. HEIN and P.~H.~E. MEIJER.
\newblock Critical magnetic fields of superconducting snte.
\newblock \emph{Phys. Rev.}, 179:\penalty0 497--511, Mar 1969.
\newblock \doi{10.1103/PhysRev.179.497}.
\newblock URL \url{https://link.aps.org/doi/10.1103/PhysRev.179.497}.

\bibitem[Steele and Hein(1953{\natexlab{a}})]{Titanium}
M.~C. Steele and R.~A. Hein.
\newblock Superconductivity of titanium.
\newblock \emph{Phys. Rev.}, 92:\penalty0 243--247, Oct 1953{\natexlab{a}}.
\newblock \doi{10.1103/PhysRev.92.243}.
\newblock URL \url{https://link.aps.org/doi/10.1103/PhysRev.92.243}.

\bibitem[Steele and Hein(1953{\natexlab{b}})]{Steele1953}
M.~C. Steele and R.~A. Hein.
\newblock Superconductivity of titanium.
\newblock \emph{Phys. Rev.}, 92:\penalty0 243--247, Oct 1953{\natexlab{b}}.
\newblock \doi{10.1103/PhysRev.92.243}.
\newblock URL \url{https://link.aps.org/doi/10.1103/PhysRev.92.243}.

\bibitem[Ambegaokar and Halperin(1969)]{ambegaokar_voltage_1969}
Vinay Ambegaokar and B.~I. Halperin.
\newblock Voltage due to thermal noise in the dc josephson effect.
\newblock \emph{Phys. Rev. Lett.}, 22:\penalty0 1364--1366, Jun 1969.
\newblock \doi{10.1103/PhysRevLett.22.1364}.
\newblock URL \url{https://link.aps.org/doi/10.1103/PhysRevLett.22.1364}.

\bibitem[Beenakker(1991)]{Beenakker1991}
C.~W.~J. Beenakker.
\newblock Theory of {C}oulomb-blockade oscillations in the conductance of a quantum dot.
\newblock \emph{Phys. Rev. B}, 44:\penalty0 1646--1656, 7 1991.
\newblock \doi{10.1103/PhysRevB.44.1646}.
\newblock URL \url{https://link.aps.org/doi/10.1103/PhysRevB.44.1646}.

\bibitem[Wunnicke(2006)]{Wunnicke2006}
Olaf Wunnicke.
\newblock Gate capacitance of back-gated nanowire field-effect transistors.
\newblock \emph{Appl. Phys. Lett.}, 89:\penalty0 083102, 8 2006.
\newblock \doi{10.1063/1.2337853}.

\bibitem[Kretinin et~al.(2010)Kretinin, Popovitz-Biro, Mahalu, and Shtrikman]{Kretinin2010}
Andrey~V. Kretinin, Ronit Popovitz-Biro, Diana Mahalu, and Hadas Shtrikman.
\newblock Multimode fabry-pérot conductance oscillations in suspended stacking-faults-free inas nanowires.
\newblock \emph{Nano Letters}, 10\penalty0 (9):\penalty0 3439--3445, 2010.
\newblock \doi{10.1021/nl101522j}.
\newblock URL \url{https://doi.org/10.1021/nl101522j}.
\newblock PMID: 20695446.

\bibitem[Liang et~al.(2001)Liang, Bockrath, Bozovic, Hafner, Tinkham, and Park]{liang_fabry_2001}
Wenjie Liang, Marc Bockrath, Dolores Bozovic, Jason~H. Hafner, M.~Tinkham, and Hongkun Park.
\newblock Fabry - {Perot} interference in a nanotube electron waveguide.
\newblock \emph{Nature}, 411\penalty0 (6838):\penalty0 665--669, June 2001.
\newblock ISSN 0028-0836, 1476-4687.
\newblock \doi{10.1038/35079517}.
\newblock URL \url{https://www.nature.com/articles/35079517}.

\bibitem[Oksanen et~al.(2014)Oksanen, Uppstu, Laitinen, Cox, Craciun, Russo, Harju, and Hakonen]{oksanen_single_2014}
Mika Oksanen, Andreas Uppstu, Antti Laitinen, Daniel~J. Cox, Monica~F. Craciun, Saverio Russo, Ari Harju, and Pertti Hakonen.
\newblock Single-mode and multimode fabry-p\'erot interference in suspended graphene.
\newblock \emph{Phys. Rev. B}, 89:\penalty0 121414, Mar 2014.
\newblock \doi{10.1103/PhysRevB.89.121414}.
\newblock URL \url{https://link.aps.org/doi/10.1103/PhysRevB.89.121414}.

\bibitem[Witmans(2025)]{witmans_2025_14959524}
Femke Witmans.
\newblock Quantum transport in snte nanowires, March 2025.
\newblock URL \url{https://doi.org/10.5281/zenodo.14959524}.

\end{thebibliography}

\newpage
\appendix
\renewcommand*{\thefigure}{S\arabic{figure}}
\setcounter{figure}{0}  
\section{Supplementary information} \label{sec:SI}

\subsection{Additional TEM results}
\begin{figure}[!ht]
\centering
\includegraphics[width=\linewidth]{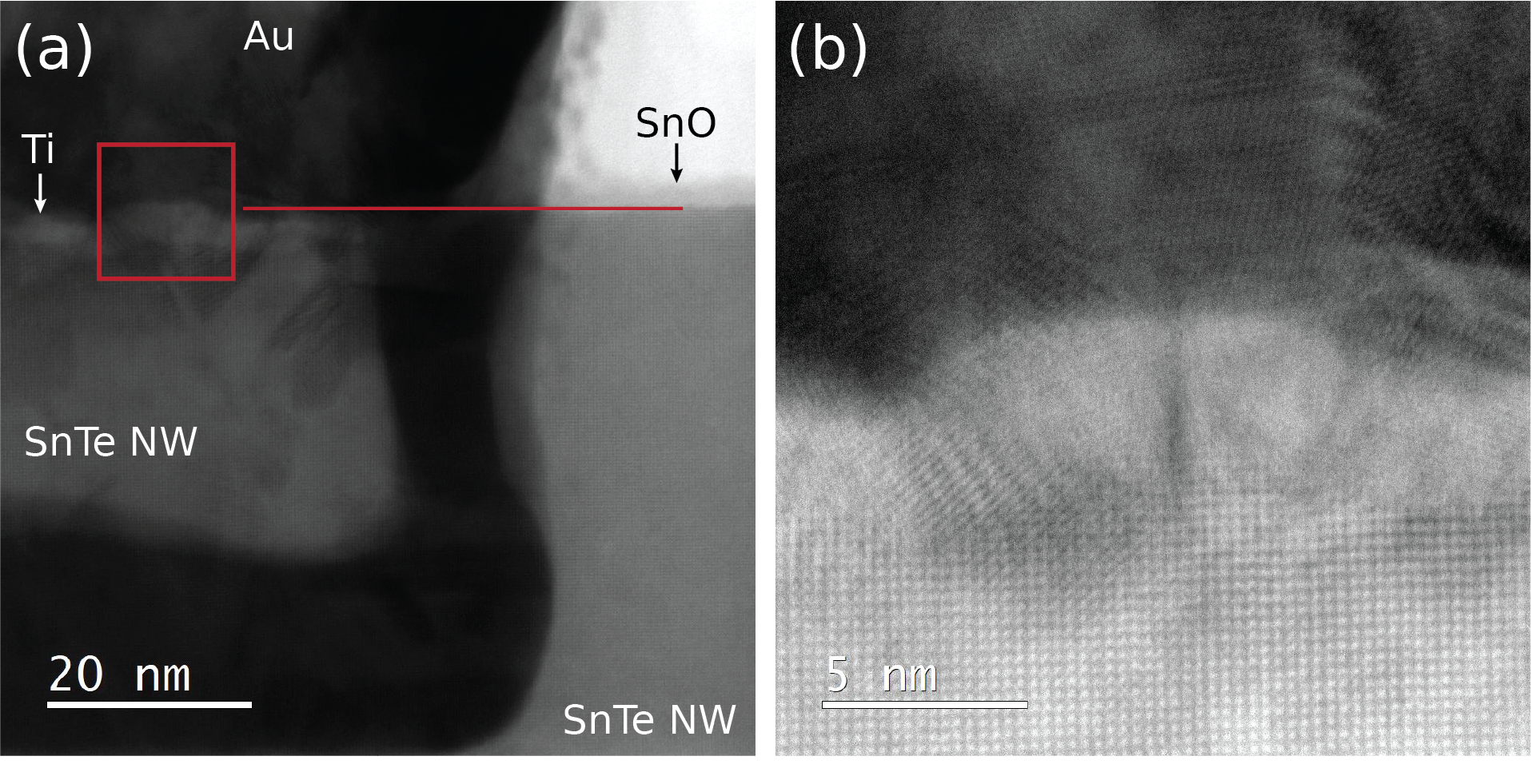}
\caption{\label{fig:BFTEM} Bright field TEM (BF-TEM) image at one of the contacts of device C1.}
\end{figure}

Figure \ref{fig:BFTEM} shows a BF-STEM image of a contact, showing the interface at the contact. The red line in Figure \ref{fig:BFTEM}a indicates the interface between the crystalline NW and the amorphous Sn oxide of the NW channel at the right side of the image. The wire underneath the contact at the left side of the image is thinner compared to the wire right of the contact. This height difference implies that the top part of the nanowire has been removed, including the native oxide layer. The height difference between the top facet of the NW at the channel side and the top part of the NW underneath the contact is large enough to conclude that the native oxide is removed. 

Because of the etching procedure, some roughness is induced in this area as well. The etching procedure entailed a sputter etch, before in-situ deposition of metal contacts. The wire is thus not exposed to air in between etching and metal deposition, following standard procedures for capping nanostructured devices after in-situ etching. In Figure \ref{fig:BFTEM}b we show the SnTe nanowire with Ti on top and above the Au capping layer. We observe that the Ti is polycrystalline with clear lattice fringes, which can be assigned to the room temperature hexagonal Ti structure (P63/mmc). It should be noted that this assignment is not unique. Similar lattice spacings can also be found in rutile TiO\textsubscript{2}.

\begin{figure}[!ht]
\centering
\includegraphics[width=0.7\linewidth]{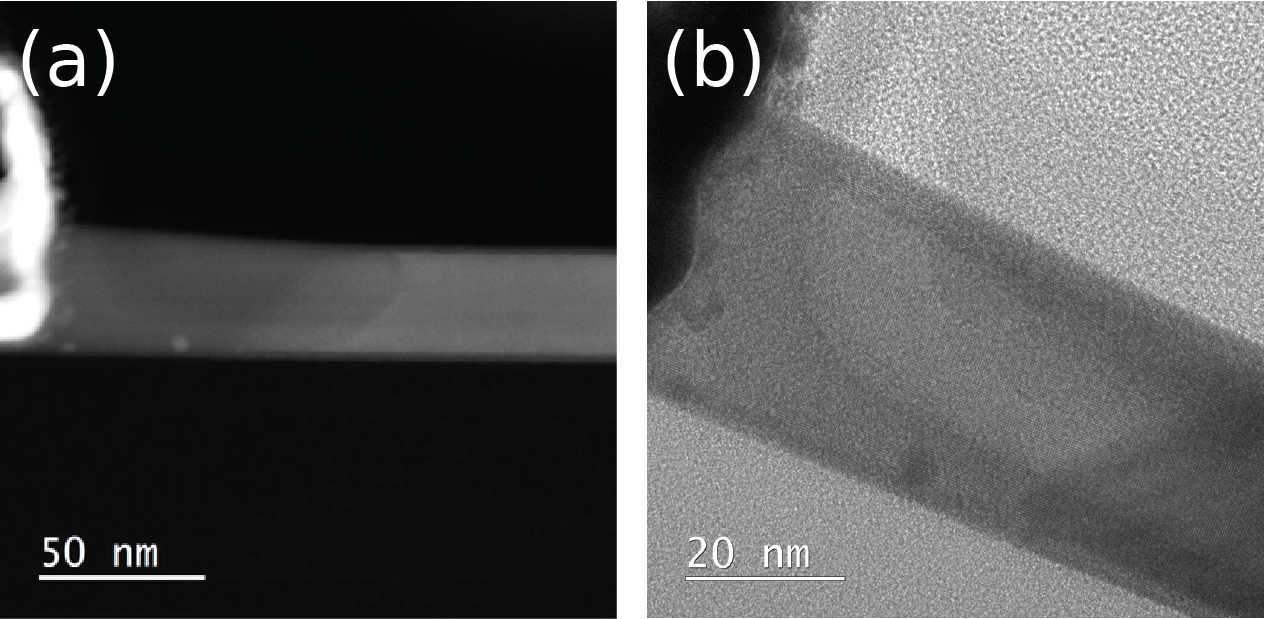}
\caption{\label{fig:HAADFTEM} HAADF-STEM (High angle annular dark field scanning TEM) images of the Sn-depleted region in device A1.}
\end{figure}

Figure \ref{fig:HAADFTEM}a,b show HAADF-STEM (High angle annular dark field scanning TEM) images of the Sn-depleted region. The reduced brightness in this image suggests that the material in this area has lower density or a lower average atomic number than the rest of the wire. 
Figure \ref{fig:HAADFTEM}b displays an HR-TEM image showing that at least part of the thickness of the nanowire is crystalline and has the same crystal structure as the rest of the NW. Possibly, within the thickness of the TEM lamella, intact SnTe and Sn-depleted material is present.

\newpage

\begin{figure}[!ht]
\centering
\includegraphics[width=0.5\linewidth]{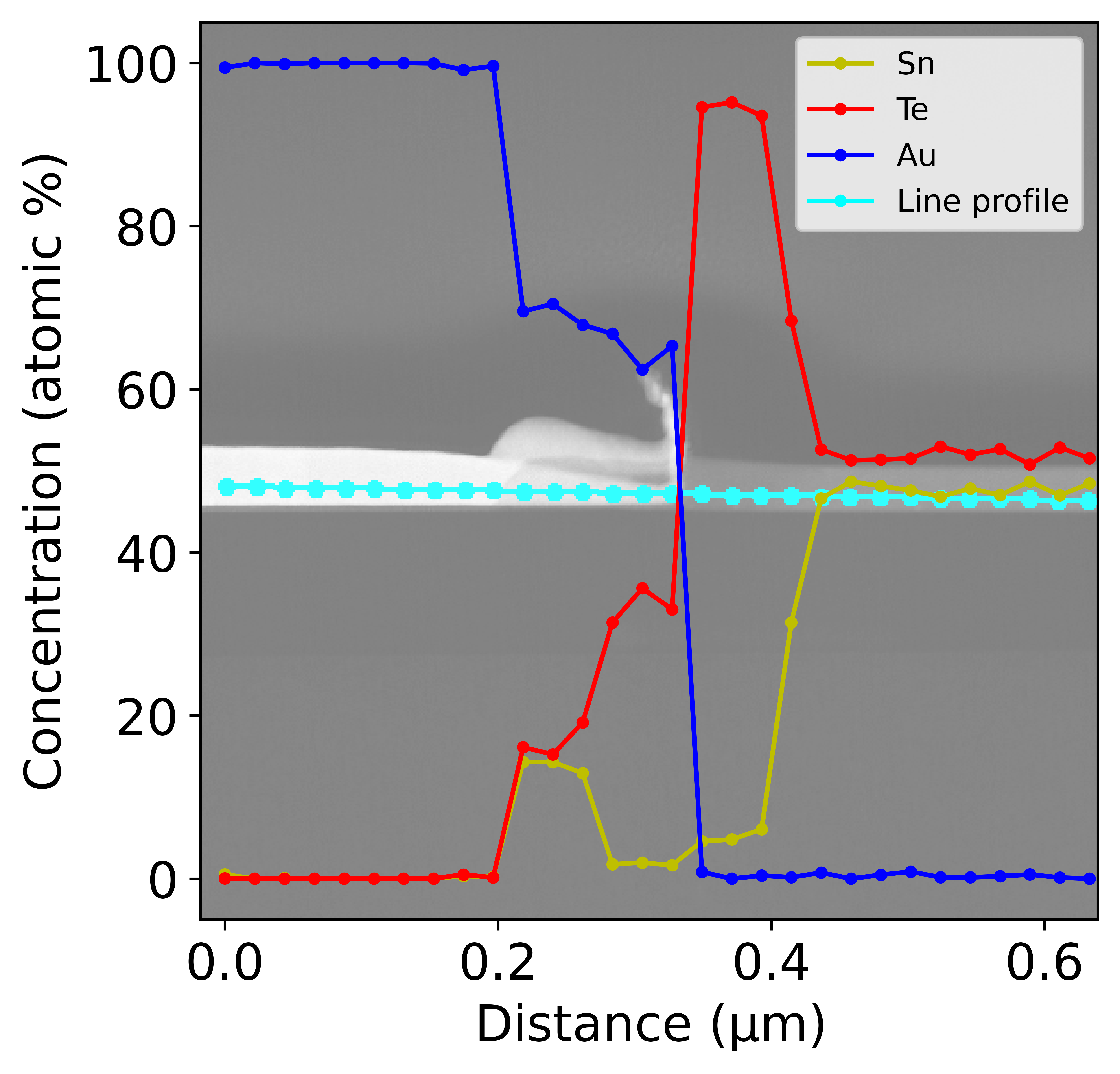}
\caption{\label{fig:EDX} EDX compositional profile, showing demixing of SnTe at the contact.}
\end{figure}

An EDX compositional profile is performed, where over a line the concentration of the elements Sn, Te and Au is scanned for. Figure \ref{fig:EDX} shows the concentration for the three elements on top of an BF-TEM image with the data point locations indicated.

The EDX mappings of the separate elements (Sn, Te, Ti, Au, O, and Si) of devices C1 and A1 are shown in the following figures. For A1 both contacts are examined separately. For C1 the FIB lamella was thinned more, resulting in the EDX mappings shown in Figure \ref{fig:EDXC1_2}. \newpage

\begin{figure}[!ht]
\centering
\includegraphics[width=0.7\linewidth]{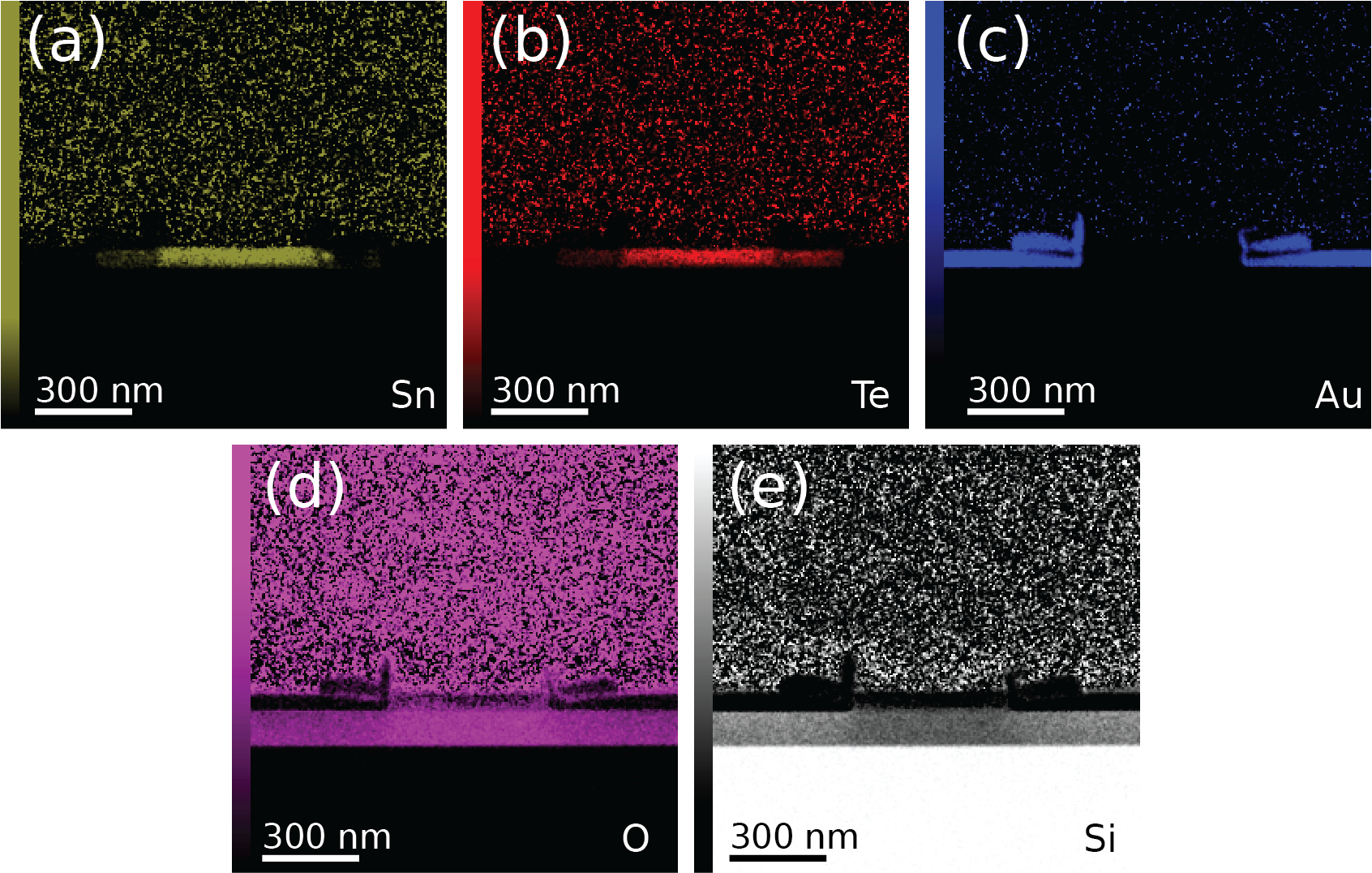}
\caption{\label{fig:EDXC1} EDX element mappings for device C1, as depicted in Figure \ref{fig:morph} as well. \textbf{(a)} Sn, \textbf{(b)} Te, \textbf{(c)} Au, \textbf{(d)} O, \textbf{(e)} Si.}
\end{figure}

We performed additional EDX measurements at the contact area, shown in figure \ref{fig:EDXC1_2}. Here we observe an O signal at the position of the Ti layer. The exact composition is difficult to determine because of reabsorption effects and because it is not clear to what extend oxidation of the TEM lamella can have influenced the O content in the Ti-layer. Ti was deposited in a vacuum chamber with a pressure lower than 1e-6 mbar, making it highly unlikely that the 3 nm of TiO\textsubscript{2} has completely oxidized. In transport measurements we see that we are able to obtain linear IV curves (see section \ref{sec:IVs}) in the low-resistance regime, indicating ohmic contacts. In 2 of the devices in the high-resistance regime, we observe quantum dot behavior, which is explained by the Sn-depleted region as elaborated on above. For the other 2 devices showing disordered behavior, the presence of a somewhat thicker layer of TiO\textsubscript{2} could be an explanation of why we observe disorder in these devices, although this remains speculation since we cannot do TEM analysis for all 11 devices. 

\begin{figure}[!ht]
\centering
\includegraphics[width=0.7\linewidth]{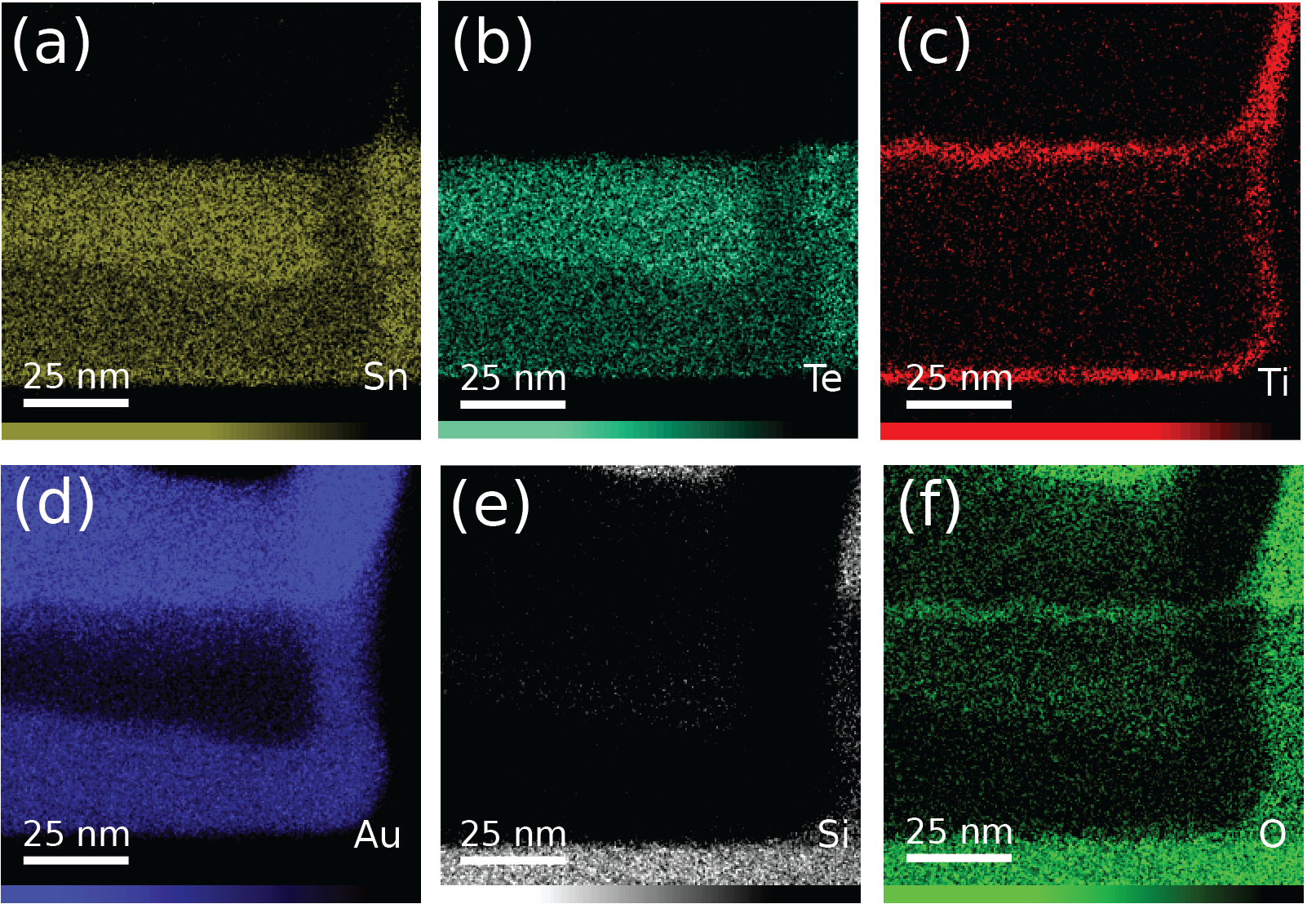}
\caption{\label{fig:EDXC1_2} EDX element mappings for device C1, zoomed in on one of the contacts.  \textbf{(a)} Sn, \textbf{(b)} Te, \textbf{(c)} Ti, \textbf{(d)} Au, \textbf{(e)} Si, \textbf{(f)} O.}
\end{figure}

\begin{figure}[!ht]
\centering
\includegraphics[width=0.7\linewidth]{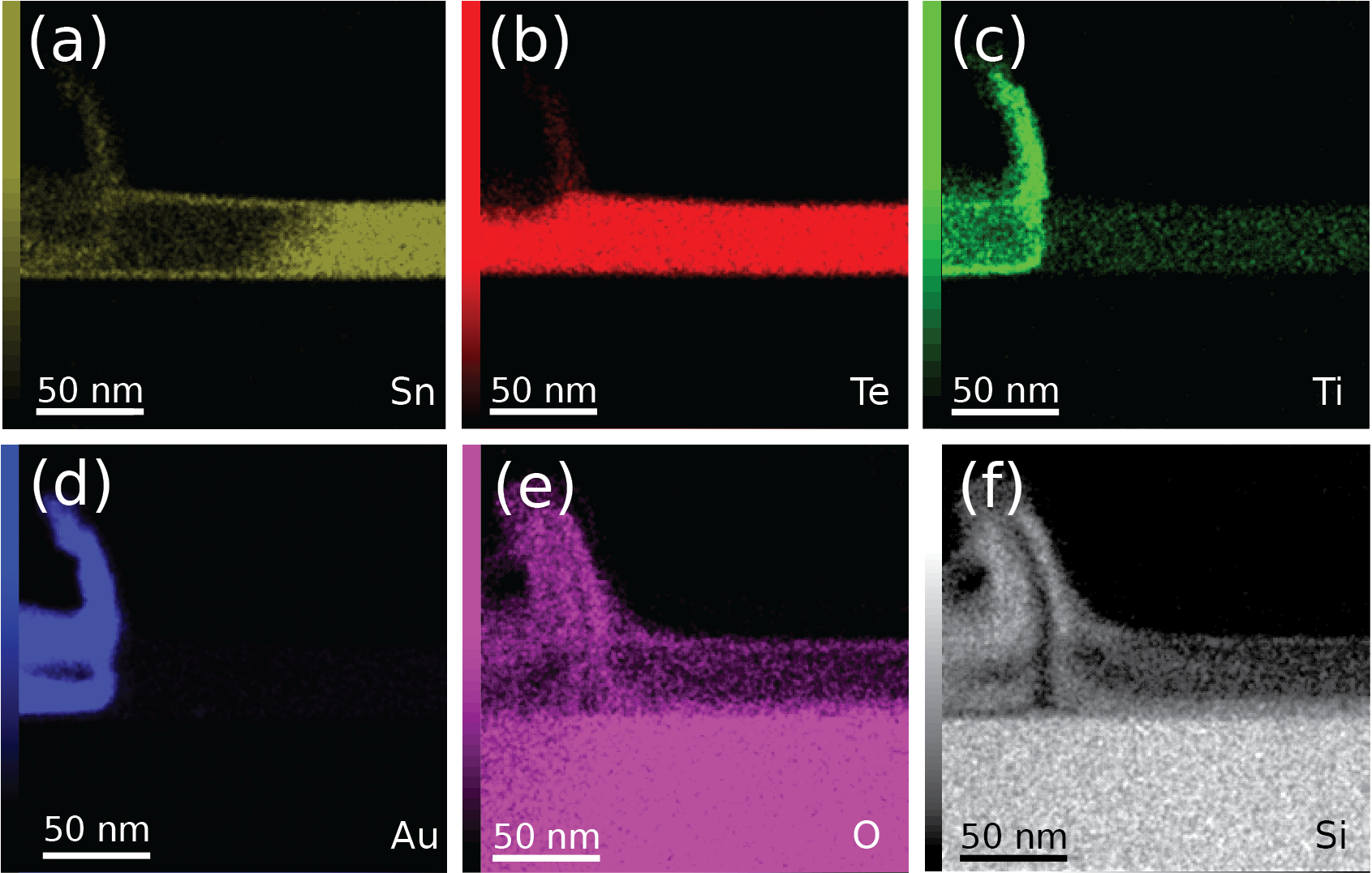}
\caption{\label{fig:EDXC5} EDX element mappings for device A1, as depicted in Figure \ref{fig:morph} as well. \textbf{(a)} Sn, \textbf{(b)} Te, \textbf{(c)} Ti, \textbf{(d)} Au, \textbf{(e)} O, \textbf{(f)} Si.}
\end{figure}

\begin{figure}[!ht]
\centering
\includegraphics[width=0.7\linewidth]{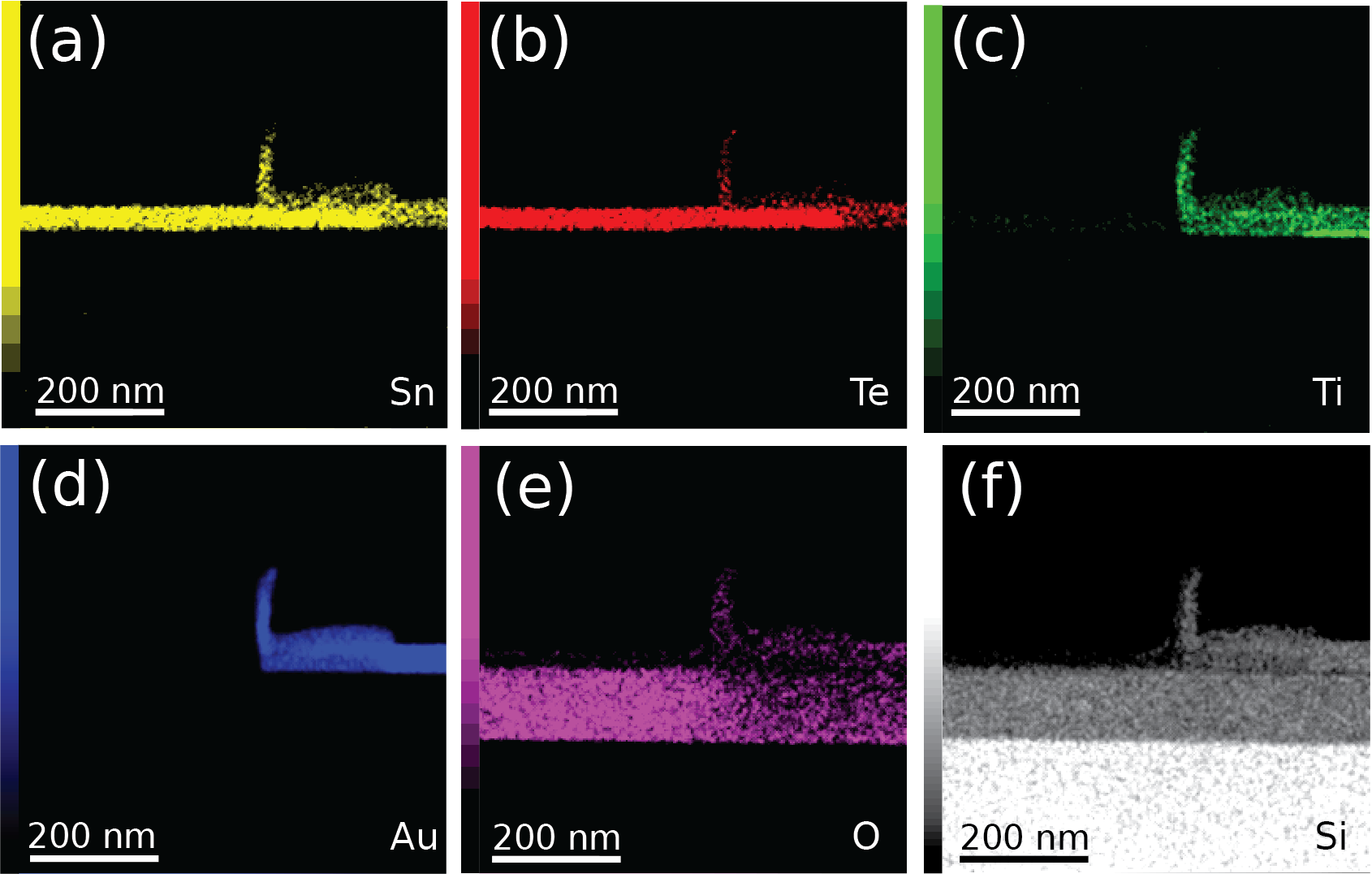}
\caption{\label{fig:EDXrightC5} EDX element mappings for device A1, other contact. \textbf{(a)} Sn, \textbf{(b)} Te, \textbf{(c)} Ti, \textbf{(d)} Au, \textbf{(e)} O, \textbf{(f)} Si.}
\end{figure}

\newpage
\subsection{Device parameters}
\begin{table}[h!]
\centering
\begin{tabular}{|c|c|c|c|}
\hline
\textbf{Device} & \textbf{Wire length (nm)} & \textbf{Channel length $L$ (nm)} & \textbf{Diameter (nm)} \\ \hhline{|=|=|=|=|}
A1     & 1136      & 791              & 37     \\ \hline
A2     & 879       & 402              & 23     \\ \hline
A3     & 868       & 597              & 22     \\ \hline
A4     & 786       & 344              & 14     \\ \hline
B1     & 700       & 323              & 45     \\ \hline
B2     & 1054      & 350              & 42     \\ \hline
B3     & 770       & 360              & 20     \\ \hline
C1     & 924       & 465              & 57     \\ \hline
C2     & 852       & 485              & 43     \\ \hline
C3     & 1152      & 710              & 40     \\ \hline
C4     & 958       & 456              & 25     \\ \hline
\end{tabular}
\end{table}

\subsection{Overview of IV curves} \label{sec:IVs}
\begin{figure}[!ht]
\centering
\includegraphics[width=\linewidth]{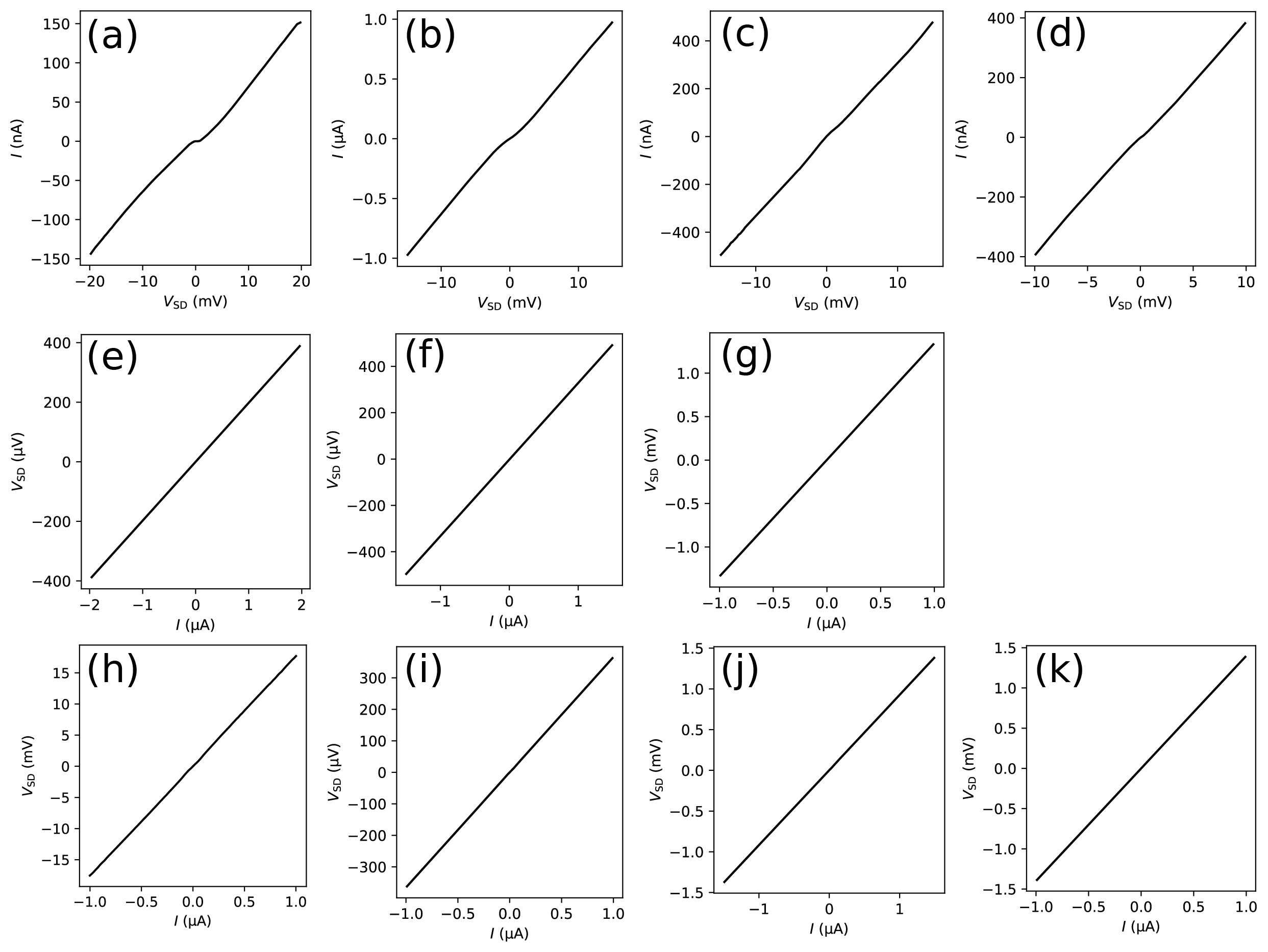}
\caption{\label{fig:IVs} IV curve of devices \textbf{(a)} A1, \textbf{(b)} A2, \textbf{(c)} A3, \textbf{(d)} A4, \textbf{(e)} B1, \textbf{(f)} B2, \textbf{(g)} B3, \textbf{(h)} C1, \textbf{(i)} C2, \textbf{(j)} C3, \textbf{(k)} C4.}
\end{figure}

In Figure \ref{fig:IVs} an overview of IV curves for each device can be found. This shows that the IV curves are linear, except for the region around 0 bias for the devices of the Semiconducting behavior type.

\subsection{More information on Hall measurements and mean free path}
Hall measurements have been performed on the same nanowire growth batch. The longitudinal voltage difference from -7 T to 7 T is approximately 3 $\Omega$. Combined with the diameter of the particular wire, this results in a carrier density of $2.8 \pm 0.1 \cdot 10^{20}$ cm\textsuperscript{-3}, when using: 
\begin{equation}
    R_H = \frac{1}{net}B.
\end{equation}

The mean free path can be extracted via the mobility: 
\begin{equation}
    l_e = \frac{\mu m v_F}{e},
\end{equation}
where the mobility is extracted from the longitudinal resistance of 68 $\Omega$. In combination with an estimation of the Fermi velocity of SnTe of $v_F = 5 \cdot10^5$ m/s [S1-S3] and effective mass of $0.1$ $m_e$, this results in a mean free path of 7 nm. 

More details on similar measurements will be included in a manuscript in preparation [S4]. 

\subsection{Parallel resistor network}
To calculate the parallel Fabry-Pérot resistor, we used the following formula: 
\begin{equation}
    \frac{1}{R_{\mathrm{total}}} = \frac{1}{R_\mathrm{bulk}} + \frac{1}{R_\mathrm{parallel}}
\end{equation}

For device B1 we take the bulk resistance as 333 $\Omega$ and the total resistance as 330 $\Omega$. For device B2 the values are 408.5 $\Omega$ and 405 $\Omega$ respectively. This results in the value of the Fabry-Pérot parallel resistor of $47 \pm 10$ k$\Omega$.

\subsection{Modelling Fabry-Pérot resonances}
\begin{figure}[h!]
\centering
\includegraphics[width=0.6\linewidth]{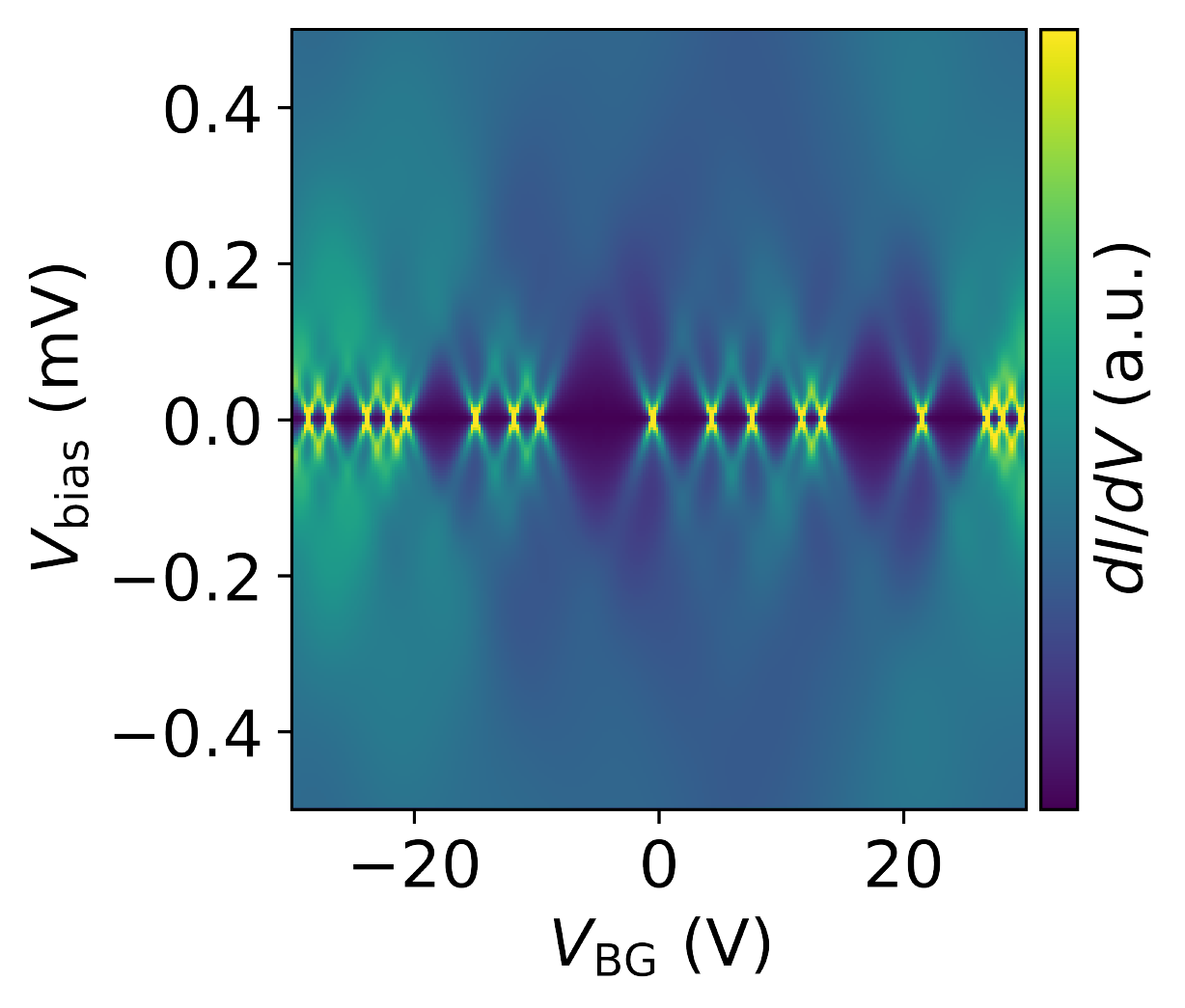}
\caption{\label{fig:FPmodel} Model of Fabry-Pérot resonances for $L=350$ $\mathrm{nm}$ and $w = 42 $ $\mathrm{nm}$ with $v_F = 5\times 10^5$ $\mathrm{m/s}$.}
\end{figure}

Equation \ref{eq:FP} is used to model Fabry-Pérot resonances combined with a broadening term, which is proportional to the bias voltage, which introduces smearing of resonances. This broadening term is a Gaussian distribution: 
\begin{equation}
    T = \frac{\Gamma}{(E-E_0)^2 + \Gamma^2},
\end{equation}
where $E$ is the energy from the applied bias voltage and back gate voltage, and $E_0$ is the energy spectrum as described by equation \ref{eq:FP}. For the model shown in figure \ref{fig:FPmodel} a range of $50 \leq n_l \leq 150$ and $0 \leq n_w \leq 15$ is chosen. For the width and length, the dimensions of the wire are chosen: 42 nm and 350 nm respectively. 

\section*{References}
[S1]  A.A. Taskin, F. Yang, S. Sasaki, K. Segawa, and Y. Ando. Topological surface transport in epitaxial SnTe thin films grown on Bi 2 Te 3. \textsl{Physical Review B
- Condensed Matter and Materials Physics}, 89(12), 2014. doi: 10.1103/PhysRevB.89.
121302. \\

\noindent [S2] Y Shi, M Wu, F Zhang, and J Feng. (111) surface states of SnTe. \textsl{PHYSICAL REVIEW B}, 90(23), December 2014. ISSN 2469-9950. doi: 10.1103/PhysRevB.90.235114.\\

\noindent [S3] Y. Tanaka, Zhi Ren, T. Sato, K. Nakayama, S. Souma, T. Takahashi, Kouji Segawa, and Yoichi Ando. Experimental realization of a topological crystalline insulator in SnTe. \textsl{Nature Physics}, 8(11):800–803, November 2012. ISSN 1745-2481. doi: 10.1038/nphys2442.\\

\noindent [S4] Manuscript in preparation

\end{document}